\documentclass [12pt] {article}
\usepackage{epsfig,afterpage}
\usepackage{colordvi,url}

\newcommand{\etal}{{\it et al.}}
\newcommand{\gev}{{\ifmmode{{{\text{GeV}}}}\else{{\mbox{GeV}}}}}

\def \Bs {B_s^0}
\def \bs {B_s^0}
\def \bd {B_d^0}
\def \bu {B_u^+}
\def \pt {p_{\rm T}}

\def \bsbar {\overline{B}_s^0}
\def \Jpsi {J/\psi}
\def \DLL {\Delta {\cal L}}
\graphicspath{{eps/}}

\begin{document}

%cbx number and authors
\begin{flushright}

LHCb-PUB-2009-015\\
S.~Stone~~\\
L.~Zhang\\
Aug. 12, 2009\\
\end{flushright}

%title
\begin{center}
{\Large\bf\boldmath Measuring the CP Violating Phase in $B_s$ Mixing Using
$B_s^0 \to J/\psi  f_0(980)$}
\end{center}
\vspace*{1 cm} \centerline{\bf Abstract} \vspace*{1cm} We show that the decay $B_s^0\to J/\psi
f_0(980)$, $f_0(980)\to\pi^+\pi^-$ can be used to measure the CP violating phase in $B_s$
mixing, $ -2\beta_s$, and estimate the sensitivity as $\pm$0.050 rad, for 2 fb$^{-1}$ of LHCb data. After adding in the
related $B_s^0\to J/\psi \eta'$, $\eta'\to \rho\gamma$ mode, the sensitivity improves to
$\pm$0.044 rad. Use of these CP eigenstates obviates the need for a transversity analysis that
must be used in the case of $B_s^0\to J/\psi \phi$ decays.

\vspace*{1 cm}
\section{Introduction}
While CP violation in $B^0$ decays has been measured unequivocally, attempts at determining the CP violating phase in $B_s$ mixing, $-2\beta_s$ \cite{Lenz} by the CDF and D0 experiments using $B_s^0\to J/\psi\phi$ decays \cite{CDF-D0}, have given
values much larger than Standard Model (SM) predictions, but with large enough errors
that the measurements are not statistically significant, even when combined \cite{Charles}. Since this
mode is not a CP eigenstate, but one involving two vector particles an angular analysis is necessary \cite{transversity}.
These analyses, however, did not allow for the possibility of there being an S-wave component
in the $\phi$ mass region, thus possibly biasing the result and surely underestimating the error \cite{stone}.

Since physics beyond the SM can or even should contribute virtual particles that
interfere in the $B_s$ mixing loop, it is important to measure the mixing phase as precisely
and in an unbiased manner as possible. It has long been known that the modes $B_s^0\to J/\psi\eta^{(')}$ are
CP eigenstates, and thus  angular analysis is not needed. All of the $\eta$ or $\eta'$ decay modes contain at least one
photon, whose reconstruction is much less efficient than charged particles in LHCb. One study has estimated the sensitivity using
$\eta'\to\rho\gamma$ \cite{Jpsietap}, but the sensitivity is of the order of a factor of two worse than that expected using
$J/\psi\phi$ mode. Other modes have also been considered \cite{other}.

In this note we describe the event selection, the backgrounds, and make an estimate of the measurement sensitivity of
$-2\beta_s$, using a heretofore not considered mode $B_s^0\to J/\psi f_0(980)$, where $f_0(980)\to \pi^+\pi^-$. The dominant Feynman diagrams for the $\phi$ and $f_0$ processes are shown in Fig.~\ref{psi_ss}. There also are possible small contributions from  penguin diagrams \cite{jpsiphi} and W-exchange {\cite{W-exchange}.

\begin{figure}[htb]
  \begin{center}
    \leavevmode
     \epsfxsize=.5 \textwidth
     \hskip 0in \epsfbox{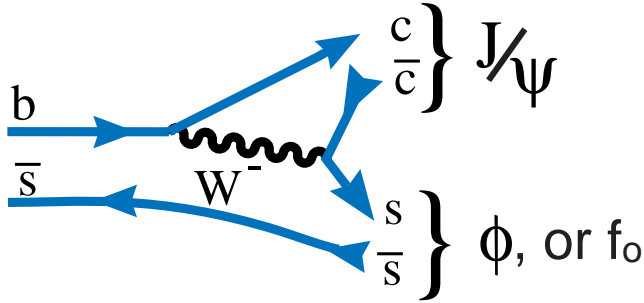}
  \end{center}
%  \vskip -0.25in
\caption{Decay diagrams for $B_s\to J/\psi \phi$, and $B_s\to J\psi f_0$.}
  \label{psi_ss}
\end{figure}

There have been several estimates of the relative widths of these decay modes.
Based on studies of hadronic $D_s^+$ decays Stone and Zhang \cite{stone} estimate that the relative width is
\begin{equation}
R_{f/\phi}\equiv\frac{\Gamma(B_s^0\to J/\psi f_0,~f_0\to \pi^+\pi^-)}{\Gamma(B_s^0\to J/\psi \phi,~\phi\to K^+K^-)}\approx 20\%~.
\end{equation}
There also is a non-$f_0$  $\pi^+\pi^-$ S-wave component that is very wide in mass, and contributes an additional 5\% of the $J/\psi\phi$ rate
using a narrow selection of $\pm$90 MeV around the $f_0$ mass that was used in the sensitivity estimate \cite{stone}.

Recently, the CLEO collaboration \cite{CLEO-f0} has reported another estimate of $R_{f/\phi}$ using measurements of the semileptonic widths of $D_s^+\to f_0 e^+\nu$ and $D_s^+\to \phi e^+\nu$ at the endpoint of the four-momentum transfer range where the phase space is maximum, as suggested by Stone and Zhang. These measurements yield
\begin{equation}
R_{f/\phi}\equiv \frac{{\frac{d\Gamma}{dq^2}}(D_s^+\to f_0(980) e^+\nu,~f_0\to\pi^+\pi^-)\mid_{q^2=0}}{{\frac{d\Gamma}{dq^2}}(D_s^+\to \phi e^+\nu,~\phi\to K^+ K^-)\mid_{q^2=0}}=(42\pm11)\%~.
\end{equation}

A note on notation, since it is clumsy to always refer to $-2\beta_s$ we set this equal to $\phi_f$ to indicate that this is the CP violating phase measured in $B_s\to J/\psi f_0$ decays, which should to an excellent approximation be the same phase as measured in $B_s\to J/\psi \phi$ decays even in the presence of new physics in the mixing amplitude.

\section{Signal Selection and Optimization}

The present study is done using a Monte Carlo simulation of the signal and specific backgrounds. For the signal, a $B_s$ is produced in
a 14 TeV proton proton collision conforming to the theoretical fragmentation function. It is allowed to decay and the quasi-stable particles then traverse the detector
where they are subject to magnetic fields, multiple scattering in material, decays and hadronic interactions \cite{DaVinci}.
To optimize the selection, we generate a signal sample, and several background samples. We use a generic
 $pp\to J/\psi~X$ sample to predict the backgrounds when we have a $J/\psi$ present in the event, though
 these events should be eliminated since the $J/\psi$ is produced at the primary vertex (PV).
 To study the background from $b$ decays we use a generic $b$ decay sample that also includes
 $b\to J/\psi~X$ decays.  However, some individual modes present specific problems, so we also
 generate them separately.  Table \ref{mc} lists the different decay channels
 that have been generated, the number of analyzed events passing the geometrical cuts, and the
 geometrical cut efficiencies.

We first generate the
events and pre-select the ones within the geometrical acceptance: for exclusive $b$ decays
we insist that the charged $B$ candidate decay tracks are below 400 mrad and larger than 10 mrad with respect
to the beam line. In the case of the inclusive $J/\psi$ sample, we insist only that both muon tracks pass
the above mentioned criteria. For the inclusive $b\overline{b}$, we only require that the one of the two
$B$ mesons be pointed within 400 mrad of the beam line.

\begin{table}[hbt]
\center
\caption{\label{mc}MC samples used in this study. }
\begin{tabular}{l|r|c}\hline
Decay & Number of events &Geometrical cut efficiency\\\hline\hline
$\bs \to \Jpsi(\mu\mu) f_0(\pi\pi)$&$60,565$& 16.4\%\\
$B_d^0 \to \Jpsi(\mu\mu) K^{*0}(K\pi)$&$2,994,516$& 17.3\%\\
$B_u^+ \to \Jpsi(\mu\mu) K^+$&$1,928,159$& 17.9\%\\
$\bs \to \Jpsi(\mu\mu) \eta^{\prime} (\rho \gamma)$&$63,319$ &16.4\%\\
$B_u^+ \to \Jpsi(\mu\mu) X$& $2,831,431$& 20.4\%\\
$B_d^0 \to \Jpsi(\mu\mu) X$ &$2,924,276$& 20.4\%\\
$\bs \to \Jpsi(\mu\mu) X$ &$720,109$&20.2\%\\
inclusive $\Jpsi(\mu\mu)$& $4,237,046$&19.7\%\\
inclusive $b\bar{b}$&$20,176,844$&43.7\%\\\hline
\end{tabular}
\end{table}

Table \ref{dc06} lists the production cross sections at 14 TeV
predicted by Pythia/EvtGen, used in our Monte Carlo
simulation.\footnote{We kept the LHC energy at 14 TeV in order
to compare with other simulations involving $\phi_f$. For 10 TeV running, the final
signal yields should be scaled by a factor of $\approx$5/7.}
\begin{table}[htb]
\center\caption{\label{dc06}Production cross sections at 14 TeV
predicted by Pythia/EvtGen, used in our  Monte Carlo
simulation. The double arrow (i.e. $\Rightarrow$) indicates the existence of possible
intermediate states, e.g. $\sigma_{pp\Rightarrow J/\psi X}$ includes $\sigma_{pp\to b\to J/\psi X}$. The ``prompt $J/\psi$'' cross section is $\sigma_{{\rm prompt}~ \Jpsi}=
\sigma_{pp\to J/\psi X} - \sigma_{pp\to b\Rightarrow J/\psi X}$ =0.266 mb.}
\begin{tabular}{l|c}\hline
cross section& mb\\\hline
$\sigma_{pp}$&102.9\\
$\sigma_{pp\to b\bar{b}}$&0.698\\
$\sigma_{pp\Rightarrow\Jpsi X}$&0.286\\
$\sigma_{pp\to b \Rightarrow\Jpsi X}$&0.0204\\\hline
\end{tabular}
\end{table}

To reconstruct the $J/\psi f_0$ ($J/\psi\to\mu^+\mu^-$, $f_0\to\pi^+\pi^-$) candidates, we first insist that we have two opposite sign muon candidates
that form a $\Jpsi$ candidate and satisfy the ``pre-selection" criteria listed in
Table~\ref{cuts}.  Then we require two additional charged tracks, that satisfy other
loose-pre-selection cuts, and are consistent with making a vertex with the two muons. The very
loose pre-selection criteria are chosen with the aim to remove as little signal as possible, and
reject large parts of the combinational background. This pre-selection is also applied to inclusive
$b\bar{b}$ and inclusive $\Jpsi$ events.

\begin{table}[htb]
\center
\caption{\label{cuts} Cut values for the $\bs \to \Jpsi f_0$ pre-selection and selection. Also
shown are the selection efficiency of each cut for the signal (absolute) and background (with
respect to ``pre-selection.") $\DLL_{i j}$ is the likelihood ratio of species $i$ relative to species $j$. The product of the $p_T$ values of $J/\psi$ daughters is
used to help eliminate any residual minimum bias background. }
\begin{tabular}{ccc@{\extracolsep{\fill}}cc}\hline
Cuts on the muons& Pre-Selection& Selection & signal
efficiency(\%)&BG efficiency(\%)\\\hline\hline
%Muon Detector& has hits &has hits &\\

$\DLL_{\mu\pi}$&$>-5$&$>-5$&97.2&-\\
$\chi^2_{\rm track}/$nDof &-&$<5$&98.9&85.9\\\hline\hline\\\hline Cuts on the $\Jpsi$& &
\\\hline\hline
Product $p_{\rm T}$ of daughters&-&$>500^2$ MeV$^2$&100&100\\
Mass Window &$\pm42$ MeV &$\pm42 $ MeV&97.9&-\\\hline\hline\\\hline

Cuts on the pions & &\\\hline\hline
$\DLL_{\pi K}$&$>-10$&$>-10$&99.2&-\\
$\DLL_{\pi\mu}$&$>-10$&$>-10$&99.9&-\\
$\chi^2_{\rm track}/$nDof &-&$<4$&95.3&65.2\\
Min IPS&-&$>3$&74.8&1.5\\\hline\hline\\\hline

Cuts on the $f_0(980)$&\\\hline\hline Sum $p_{\rm T}$ of
daughters&-&$>900$ MeV& 98.6&57.6\\
Mass Window &$\pm500$ MeV &$\pm90 $ MeV&-&-\\\hline\hline\\\hline

Cuts on the $\bs$ &&\\\hline\hline
Max IPS&- &$<5$&99.7&67.5\\
vertex fit $\chi^2$&$<50$&$<17$&88.1&51.8\\
$\cos \theta_p$ &-&$>0.99993$&80.4&4.8\\
Mass Window & $\pm300$ MeV&$\pm 50$ MeV&98.5&-\\\hline\hline
\end{tabular}
\end{table}

\subsection{\boldmath $\Jpsi\to \mu^+\mu^-$ selection}
Muon candidates are selected by requiring the track has hits in the Muon Chambers and satisfies the
identification criterion that requires
global $\Delta {\cal L} _{\mu\pi}>-5$, and has $\chi^2_{\rm track}$ per \#~of~degrees~of~freedom
(nDOF) of the Kalman fit of the track $<$~5 (Fig.~\ref{dis-muon} (d)). Specifically, $\Delta
{\cal L} _{p_1p_2}$ is the difference in the log of the likelihood between the hypothesis that this particle is of
type of $p_1$ rather than $p_2$ (equivalent to the likelihood ratio).
These cuts were studied in Ref. \cite{jpsiphi_sel} using minimum
bias events and are aimed at rejecting hadrons misidentified as muons due to random combinations of
spurious hits in the Muon Chambers. The distributions of transverse momentum ($p_{\rm T}$) for
$\mu^+$ vs. $\mu^-$, the minimum impact parameter significance (IPS) with respect to each
primary vertex,\footnote{The Impact
Parameter Significance is defined as the track impact parameter with respect to a primary vertex
divided by its error: IPS $\equiv~(\frac{\rm IP}{\sigma_{\rm IP}})$. At the luminosity of $2\times 10^{-32}$ cm$^{-2}$s$^{-1}$ about 35\% of the events have more than one primary vertex, and we choose the smallest IPS.} $\chi^2_{\rm track}/{\rm nDOF}$ and momentum of
muon candidates are shown in Fig.~\ref{dis-muon}(a-e) from the signal and inclusive $b\bar{b}$
MC after ``pre-selection."

\begin{figure}[htbp]
\center
\includegraphics[width=0.43\textwidth]{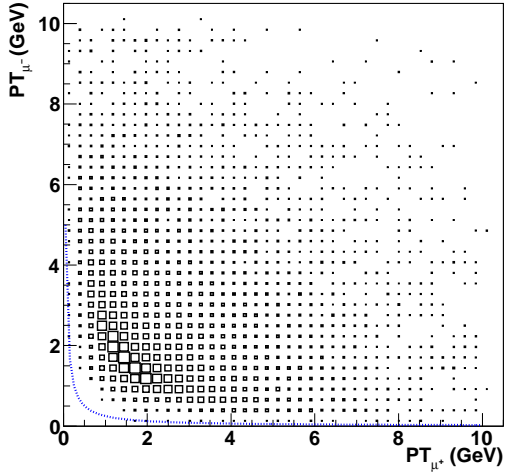}%
\includegraphics[width=0.43\textwidth]{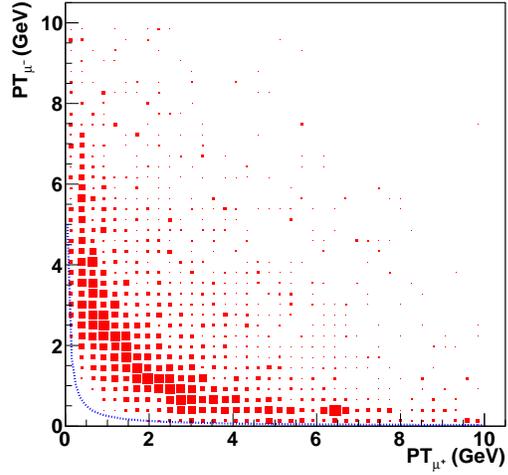}\break
\hbox to 0.43\textwidth{\hfil (a) $\pt(\mu^+)$ vs $\pt(\mu^-)$ -signal  \hfil}%
\hbox to 0.43\textwidth{\hfil (b) $\pt(\mu^+)$ vs $\pt(\mu^-)$
-background
 \hfil}\break
\includegraphics[width=0.43\textwidth]{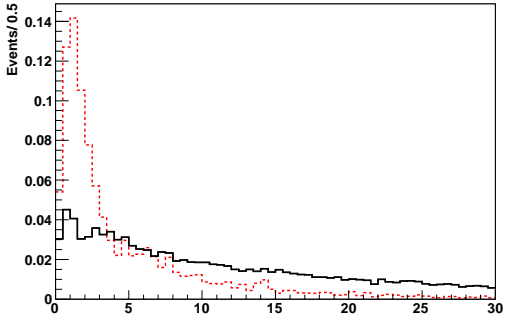}%
\includegraphics[width=0.43\textwidth]{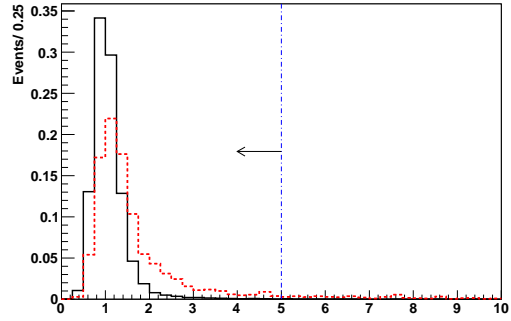}\break
\hbox to 0.43\textwidth{\hfil (c) Minimum IPS of muons  \hfil}%
\hbox to 0.43\textwidth{\hfil (d) $\chi^2_{\rm track}/{\rm nDOF}$ of
muons
 \hfil}\break
\includegraphics[width=0.43\textwidth]{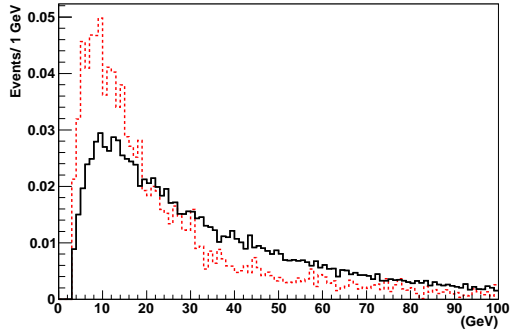}\break
\hbox to 0.43\textwidth{\hfil (e) Momentum of muons  \hfil}
%\hbox to 0.43\textwidth{\hfil (d) track fit $\chi^2$/d.o.f. of muons\hfil}
\break

\caption{\label{dis-muon}The distributions of the signal (black
solid) and $b\bar{b}$ background (red dashed) for muon candidates
from $\Jpsi$. The curves in (a) and (b) show the product of $\pt(\mu^+)\cdot\pt(\mu^-)=500^2$ MeV$^2$,
used to eliminate minimum bias events in the preselection. The final cut in (d) is indicated by vertical blue dotted
lines.}
\end{figure}

We then combine two opposite sign muon candidates to form a $\Jpsi$ candidate.
Fig.~\ref{dis-jpsi} shows the signal and $b\overline{b}$ distributions for IPS,
the flight distance significance (the distance from PV to
reconstruction vertex divided by its error), the vertex fit $\chi^2$ and invariant mass of
$\Jpsi$ candidates from the signal and inclusive $b\bar{b}$ MC after ``pre-selection."
We retain those events having $|m_{\mu\mu}-m_{\Jpsi}|<42$ MeV corresponding to about a $\pm 3\sigma$ interval
(Fig. \ref{dis-jpsi} (d)). These selections result in a relatively clean $\Jpsi$ sample.

\begin{figure}[htbp]
\center
\includegraphics[width=0.43\textwidth]{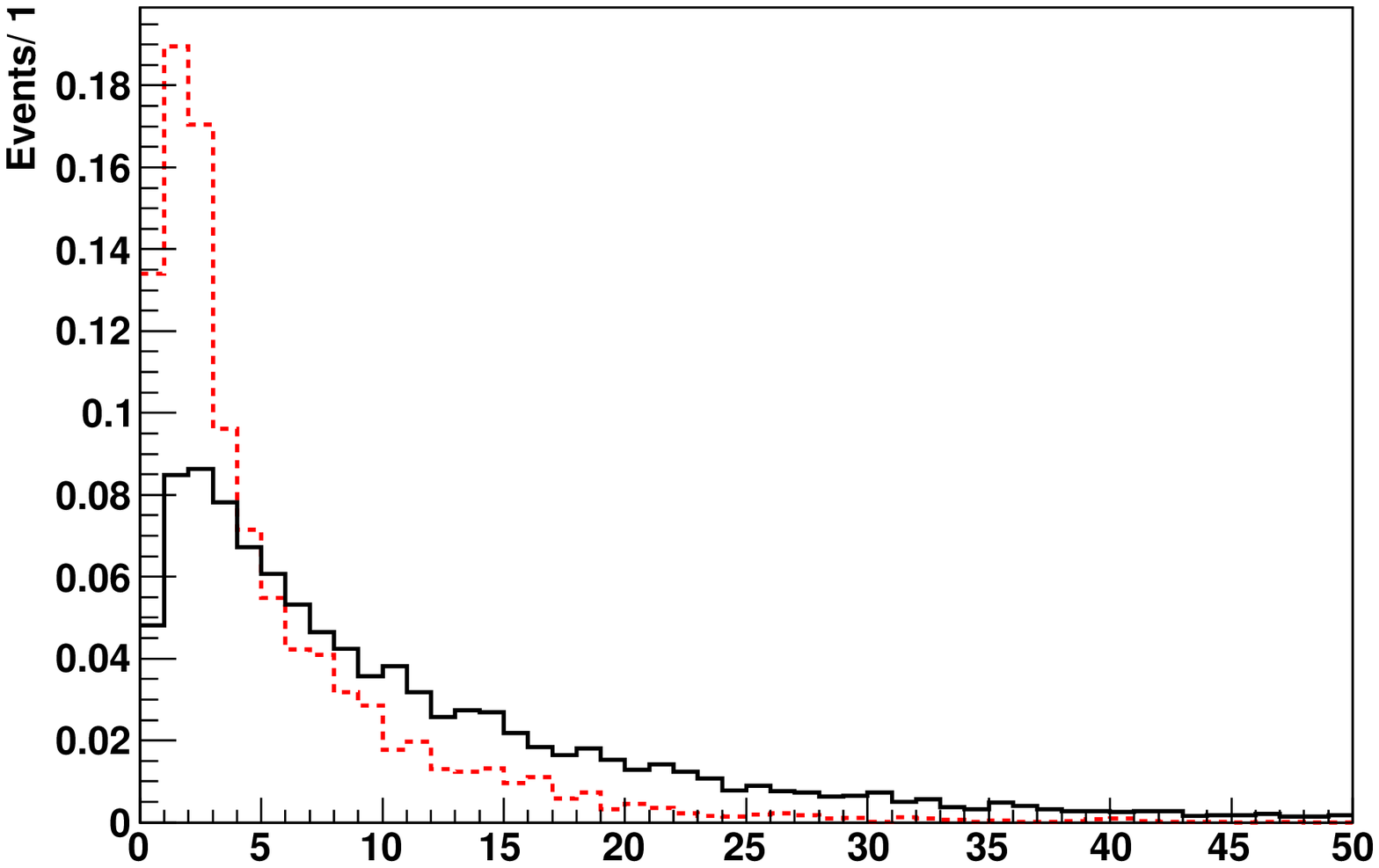}%
\includegraphics[width=0.43\textwidth]{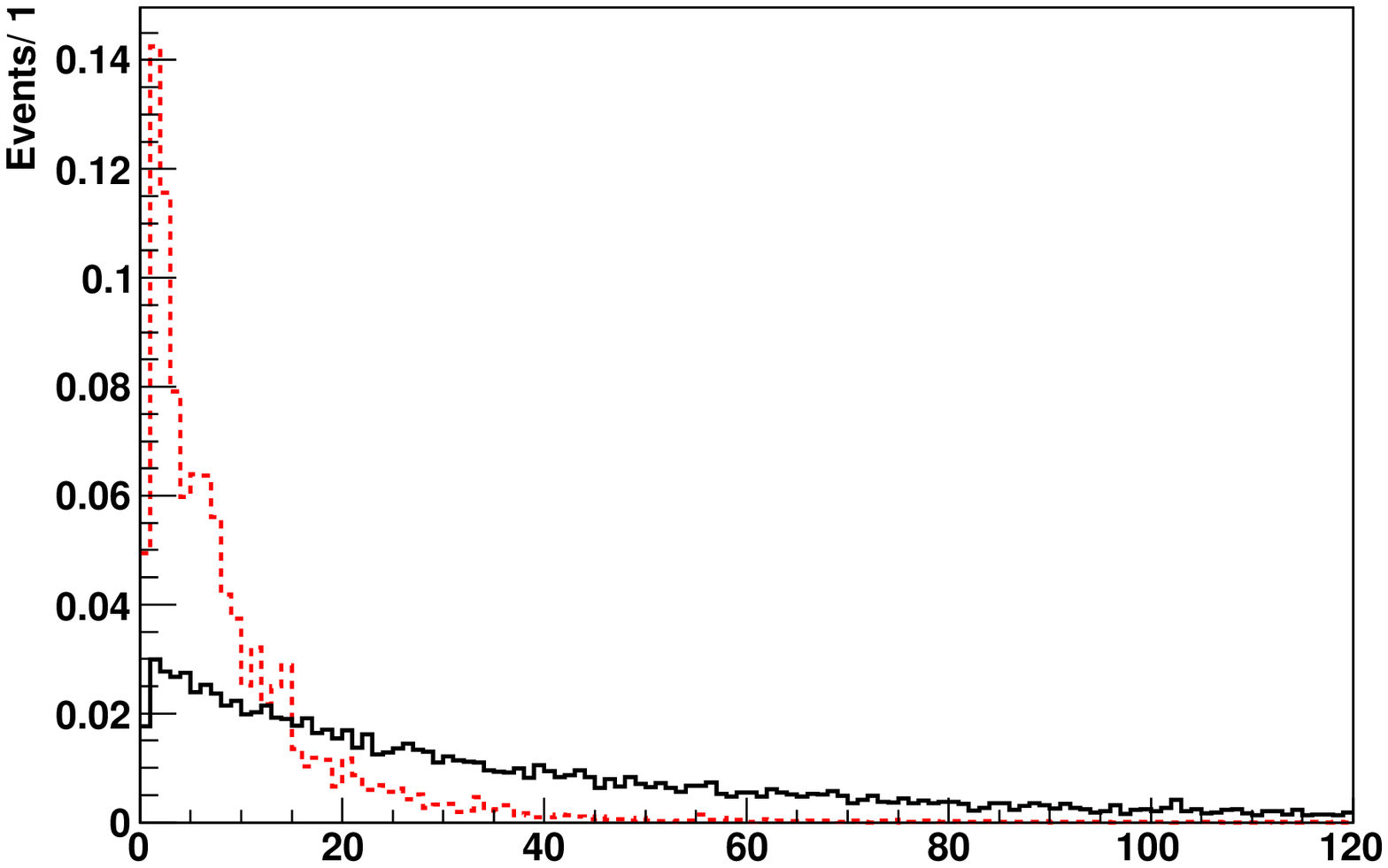}\break
\hbox to 0.43\textwidth{\hfil (a) $\Jpsi$ minimum IPS   \hfil}%
\hbox to 0.43\textwidth{\hfil (b) $\Jpsi$ flight significance
 \hfil}\break
\includegraphics[width=0.43\textwidth]{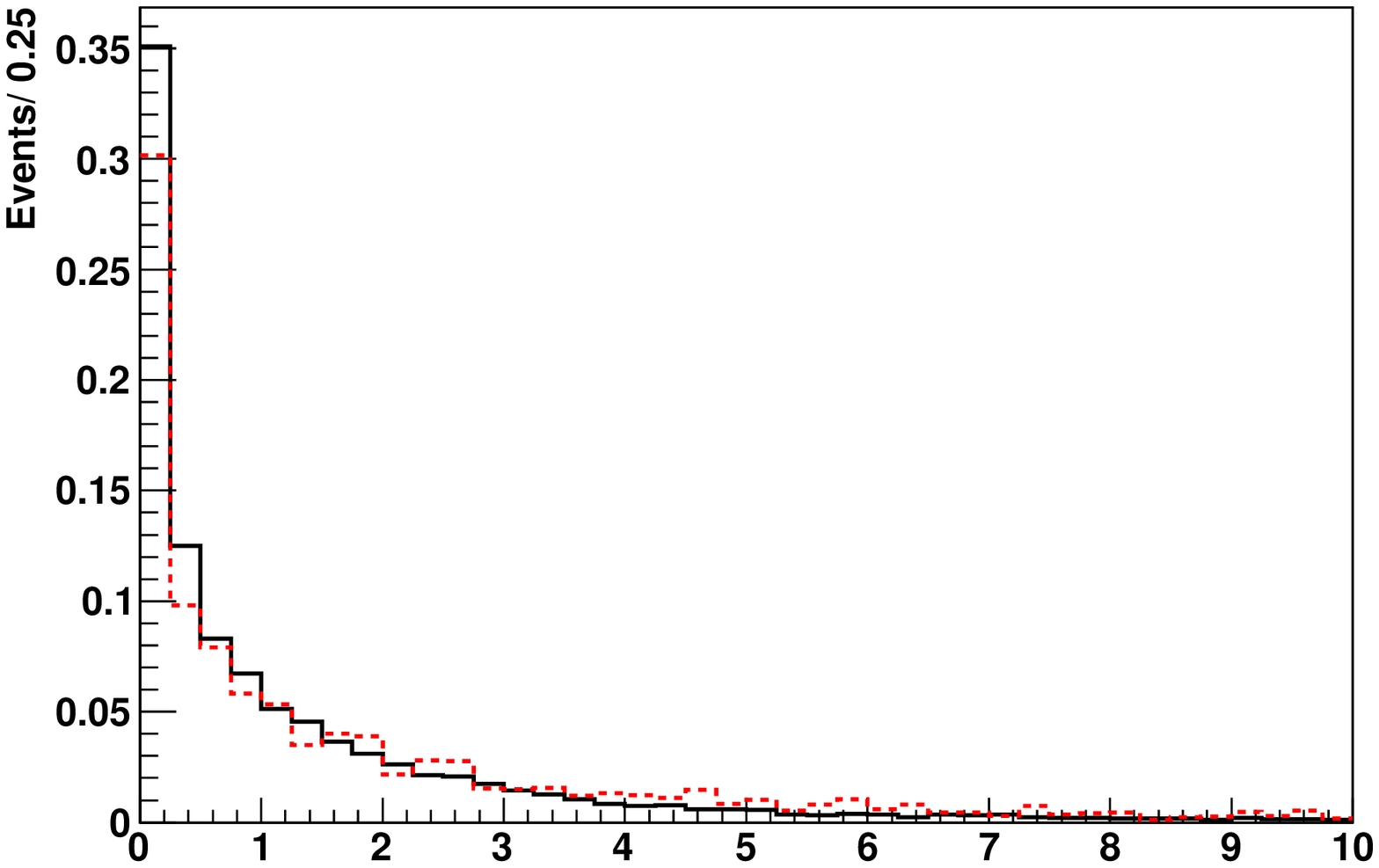}%
\includegraphics[width=0.43\textwidth]{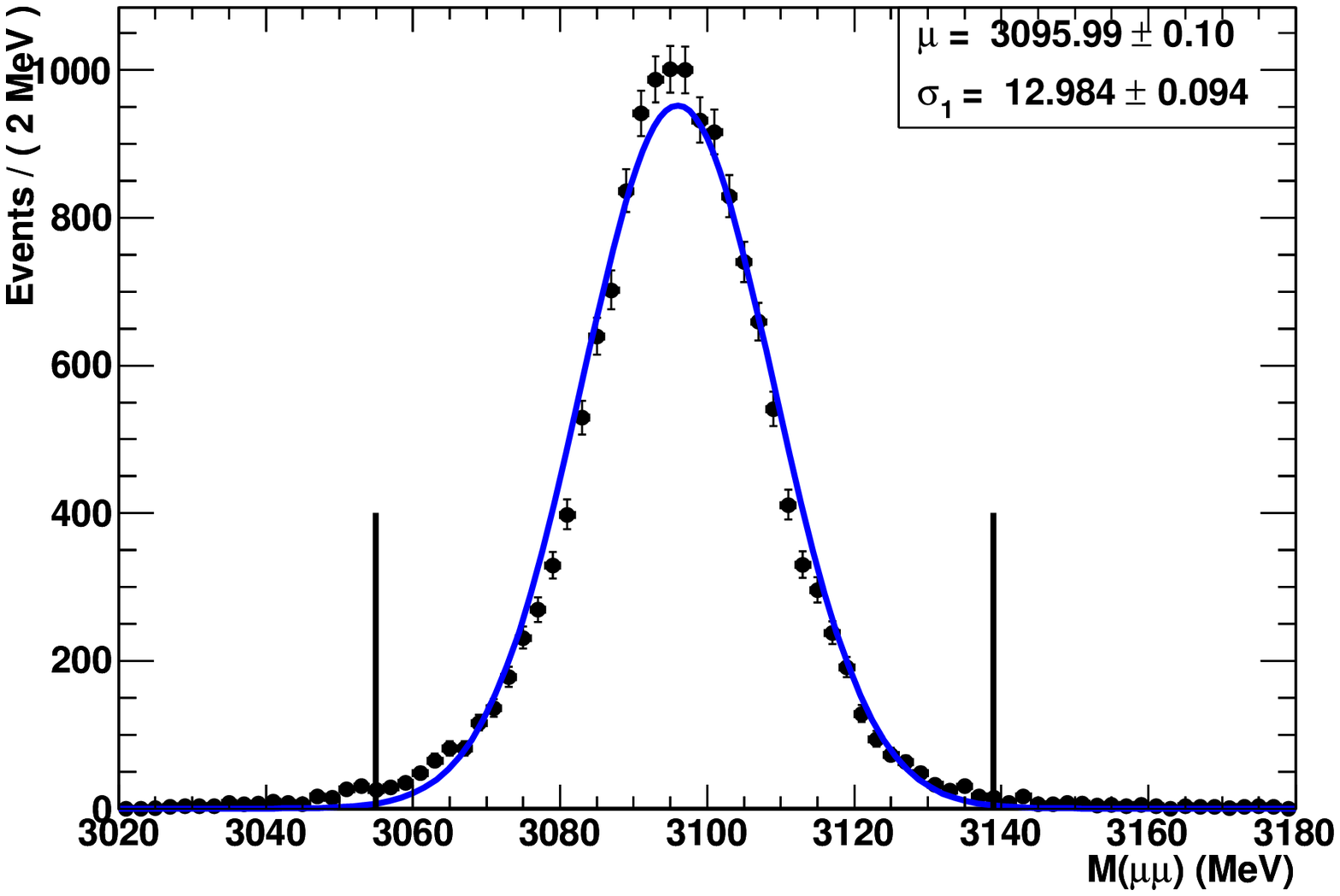}\break
\hbox to 0.43\textwidth{\hfil (c) $\Jpsi$ vertex $\chi^2$ \hfil}%
\hbox to 0.43\textwidth{\hfil (d) $\Jpsi$ mass \hfil}\break

\caption{\label{dis-jpsi}(a)-(c): The distributions of signal (black
solid) and $b\bar{b}$ background (red dashed) for $\Jpsi$ candidates. (Numbers of events in the $B$ and $S$ distributions are equal.) (d) The $J/\psi$ mass distribution for signal, fit to a single Gaussian, with the selection cut indicated by
vertical lines. }
\end{figure}

\subsection{\boldmath $f_0(980)\to \pi^+\pi^-$ selection}
To select pion candidates we veto tracks identified as kaons with $\DLL_{K\pi}>10$ or muons with
$\DLL_{\mu\pi}>10$. The veto has $>99\%$ efficiency for signal. Fig.~\ref{dis-pion} shows the
distributions of $p_{\rm T}$ for $\pi^+$ vs. $\pi^-$, the minimum IPS with respect to each
primary vertex, $\chi^2_{\rm track}/{\rm nDOF}$ and momentum of pion candidates from the signal
and inclusive $b\bar{b}$ MC after ``pre-selection."

We then combine two opposite sign pion candidates to form a $f_0(980)$ candidate. Fig.~\ref{dis-f0} shows the minimum IPS, flight significance, and vertex $\chi^2$ distributions for $f_0(980)$ candidates from the signal and inclusive
$b\bar{b}$ samples after ``pre-selection."

\begin{figure}[htbp]
\center
\includegraphics[width=0.43\textwidth]{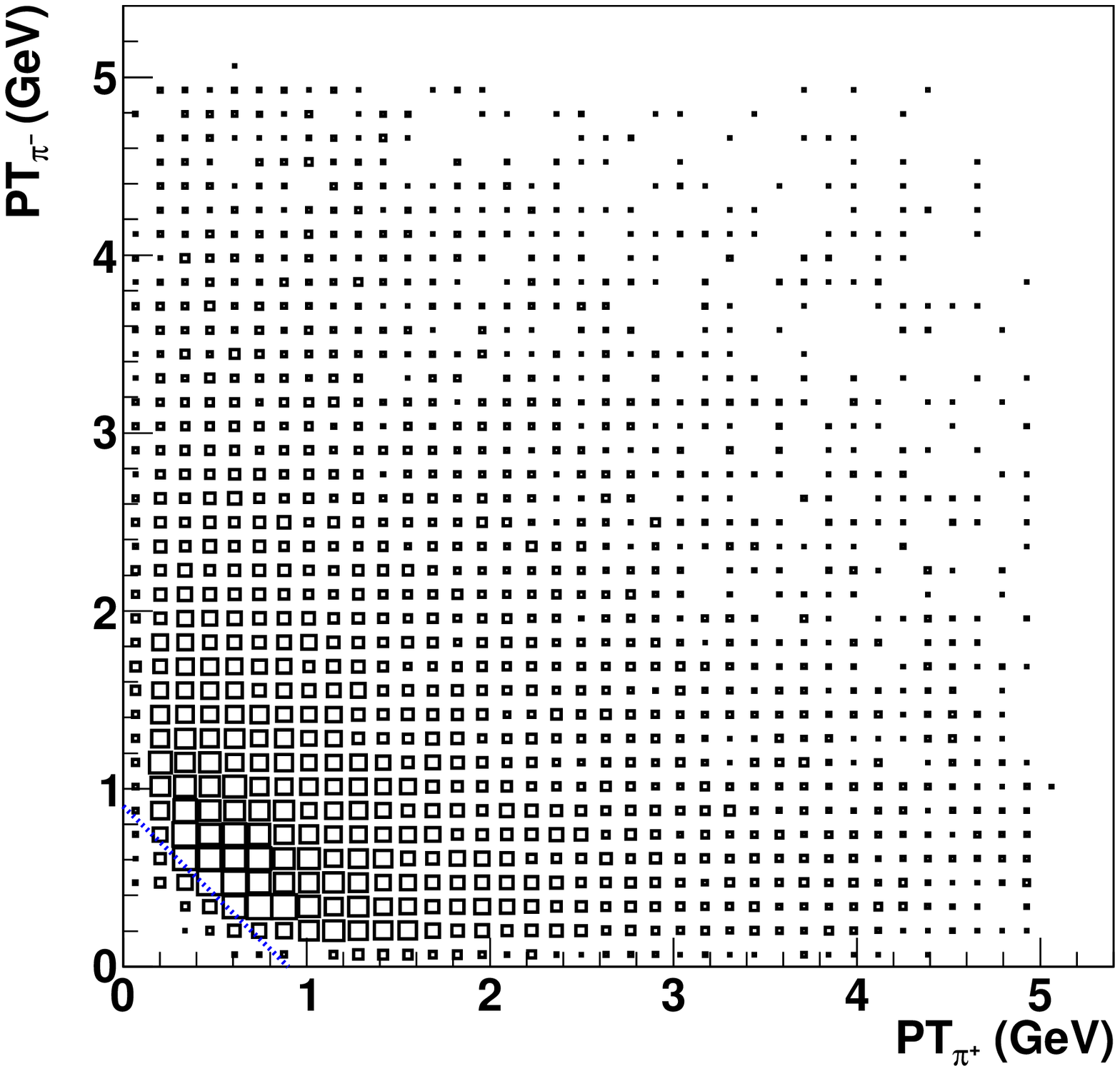}%
\includegraphics[width=0.43\textwidth]{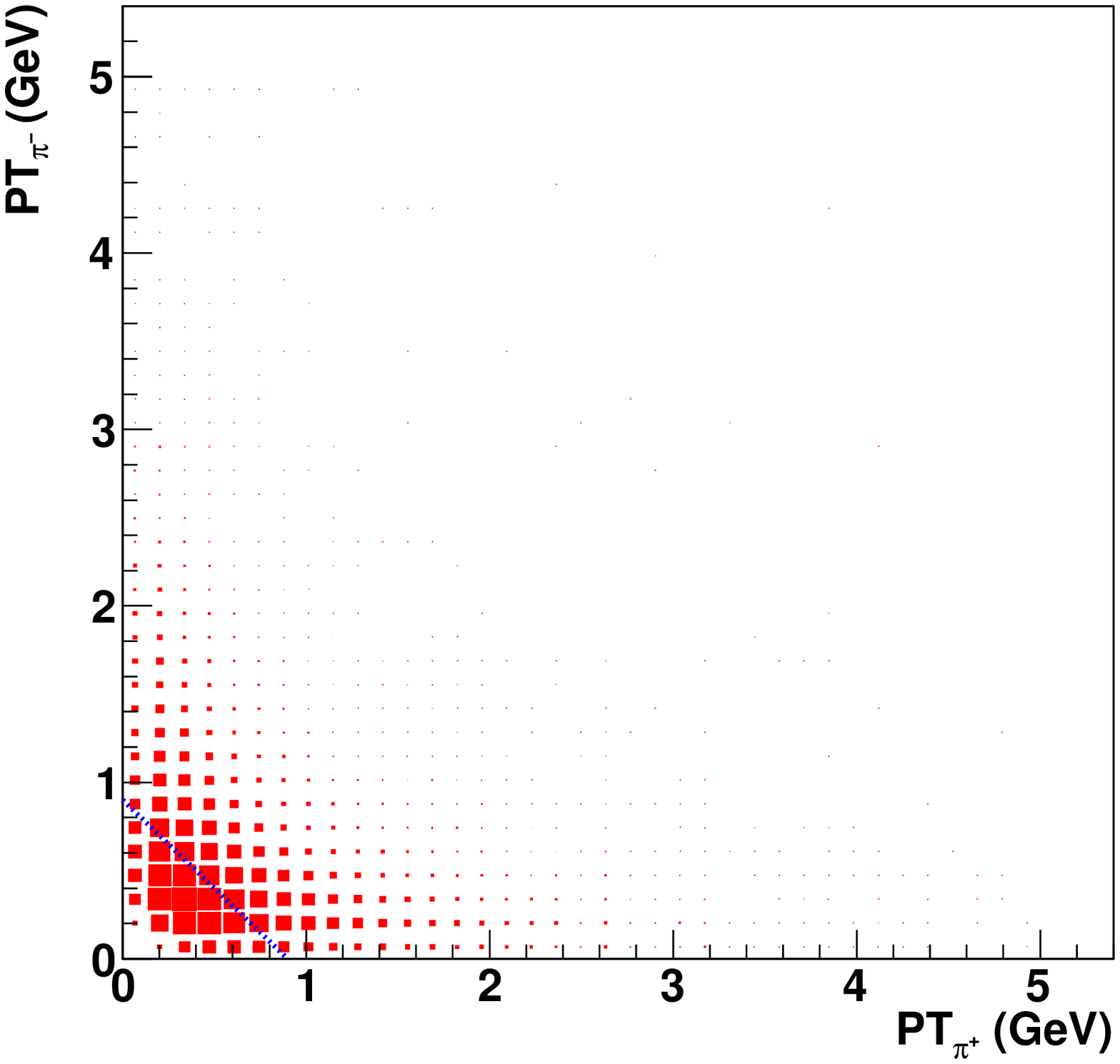}\break
\hbox to 0.43\textwidth{\hfil (a) $\pt(\pi^+)$ vs $\pt(\pi^-)$ -signal  \hfil}%
\hbox to 0.43\textwidth{\hfil (b) $\pt(\pi^+)$ vs $\pt(\pi^-)$
-background
 \hfil}\break
\includegraphics[width=0.43\textwidth]{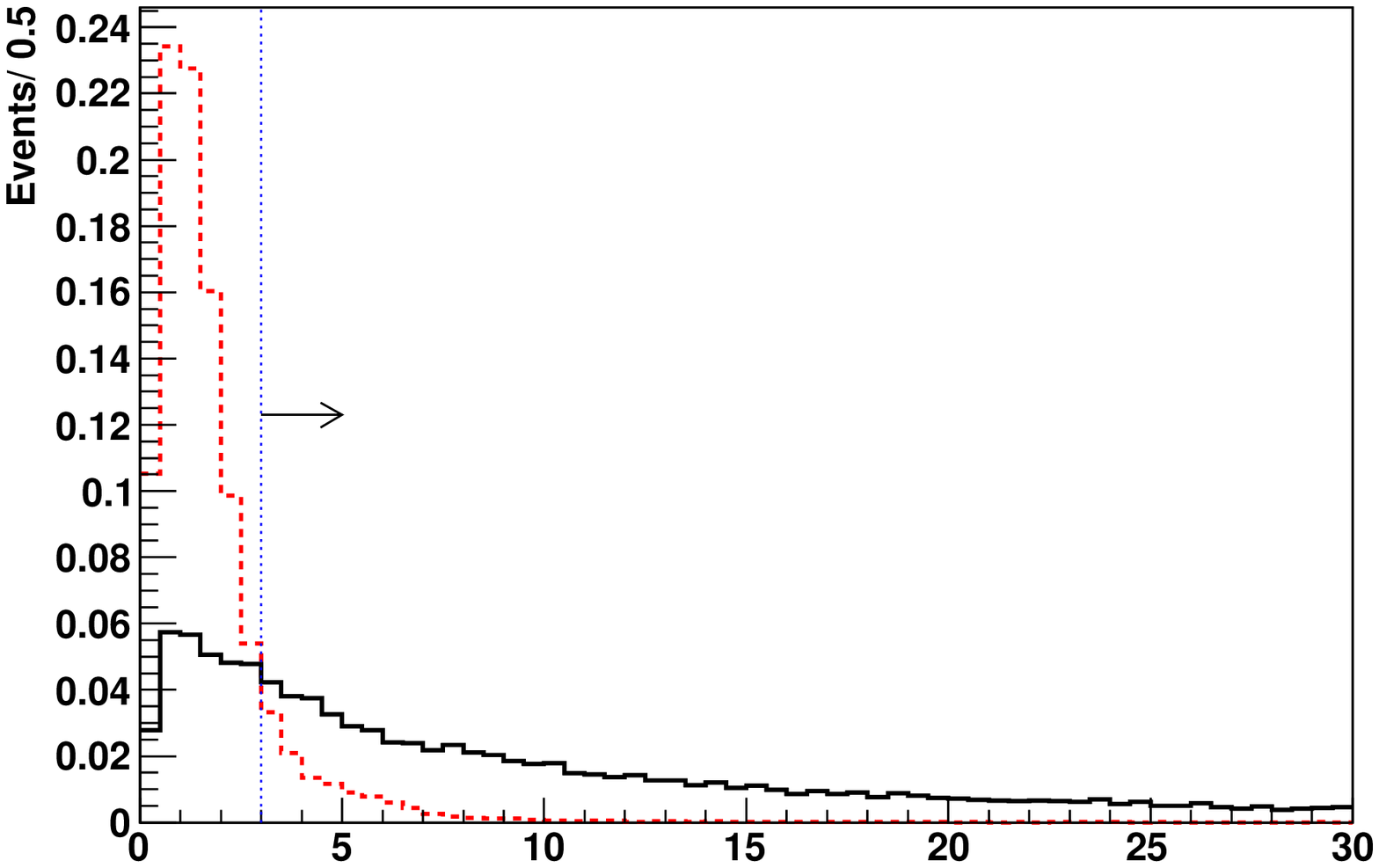}%
\includegraphics[width=0.43\textwidth]{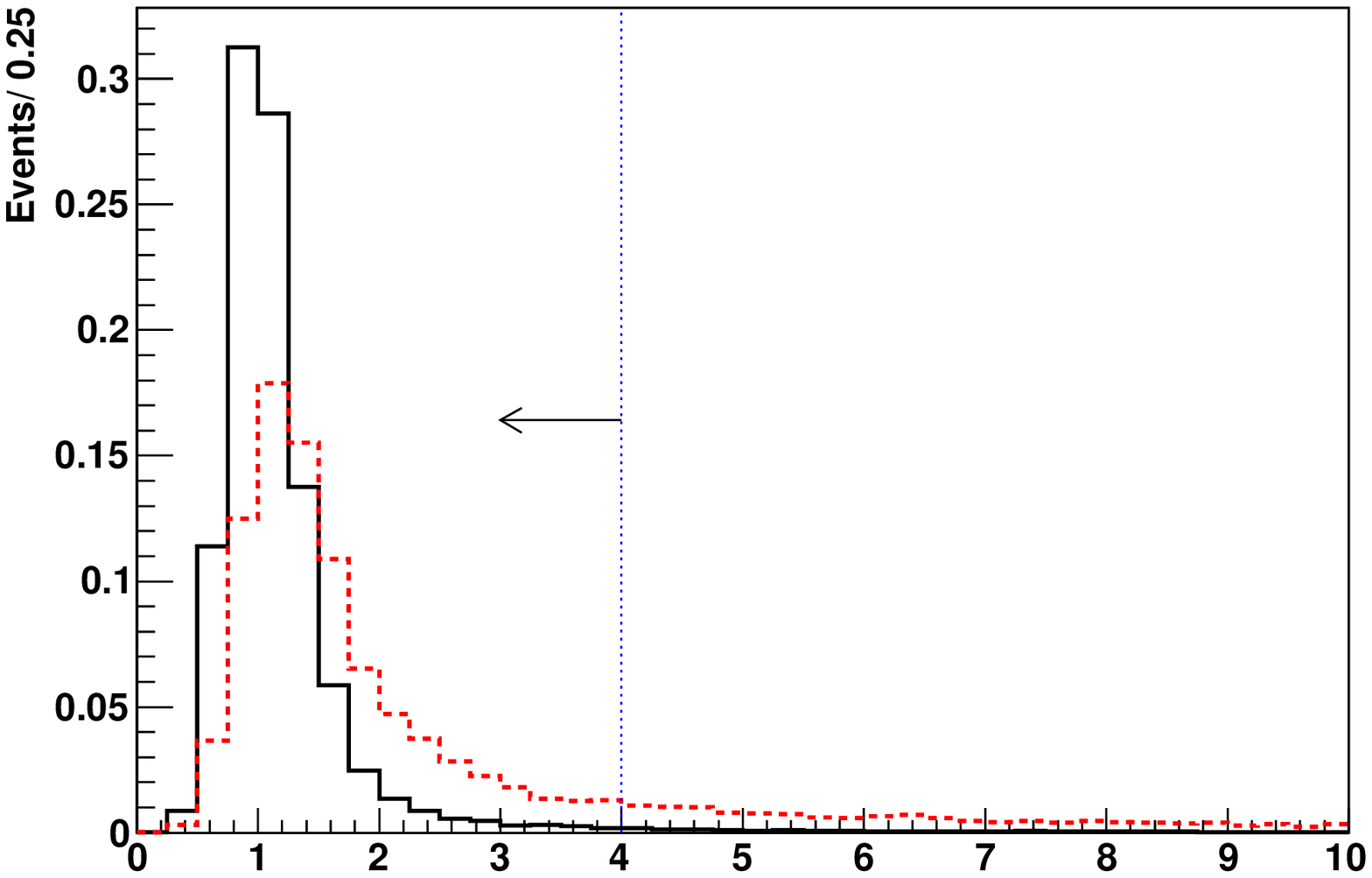}\break
\hbox to 0.43\textwidth{\hfil (c) Minimum IPS of pions  \hfil}%
\hbox to 0.43\textwidth{\hfil (d)  $\chi^2_{\rm track}$/nDOF of
pions
 \hfil}\break
 \includegraphics[width=0.43\textwidth]{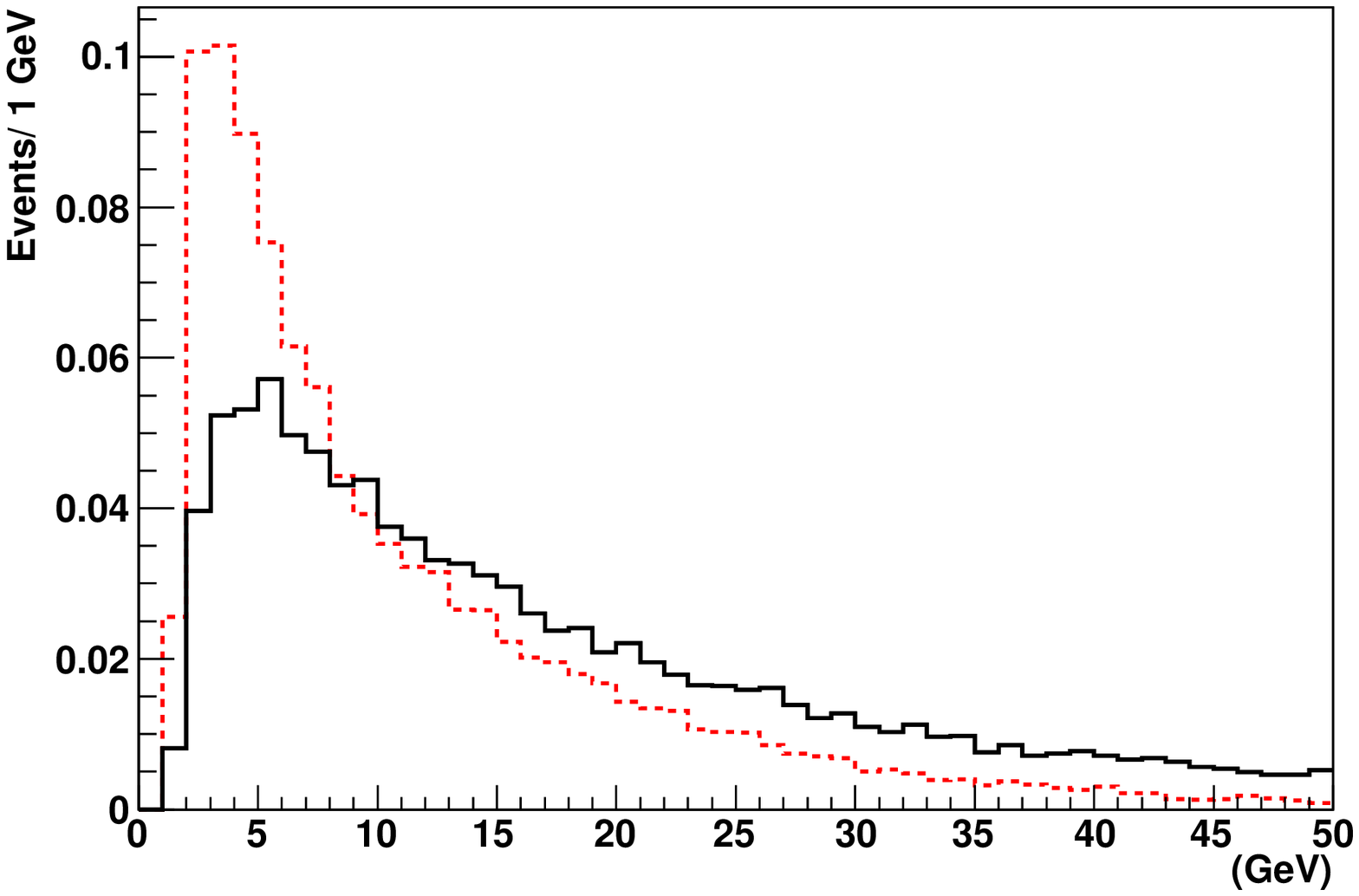}\break
\hbox to 0.43\textwidth{\hfil (e) Momentum of pions  \hfil}
%\hbox to 0.43\textwidth{\hfil (d) track fit $\chi^2$/d.o.f. of muons\hfil}
\break

\caption{\label{dis-pion}The distributions of signal (black
solid) and $b\bar{b}$ background (red dashed) for pion candidates
from $f_0$. The final cuts in (a)-(d) are indicated by blue dotted
lines. (The number of events in the $B$ and $S$ distributions are equal.)}
\end{figure}

\begin{figure}[htbp]
\center
\includegraphics[width=0.43\textwidth]{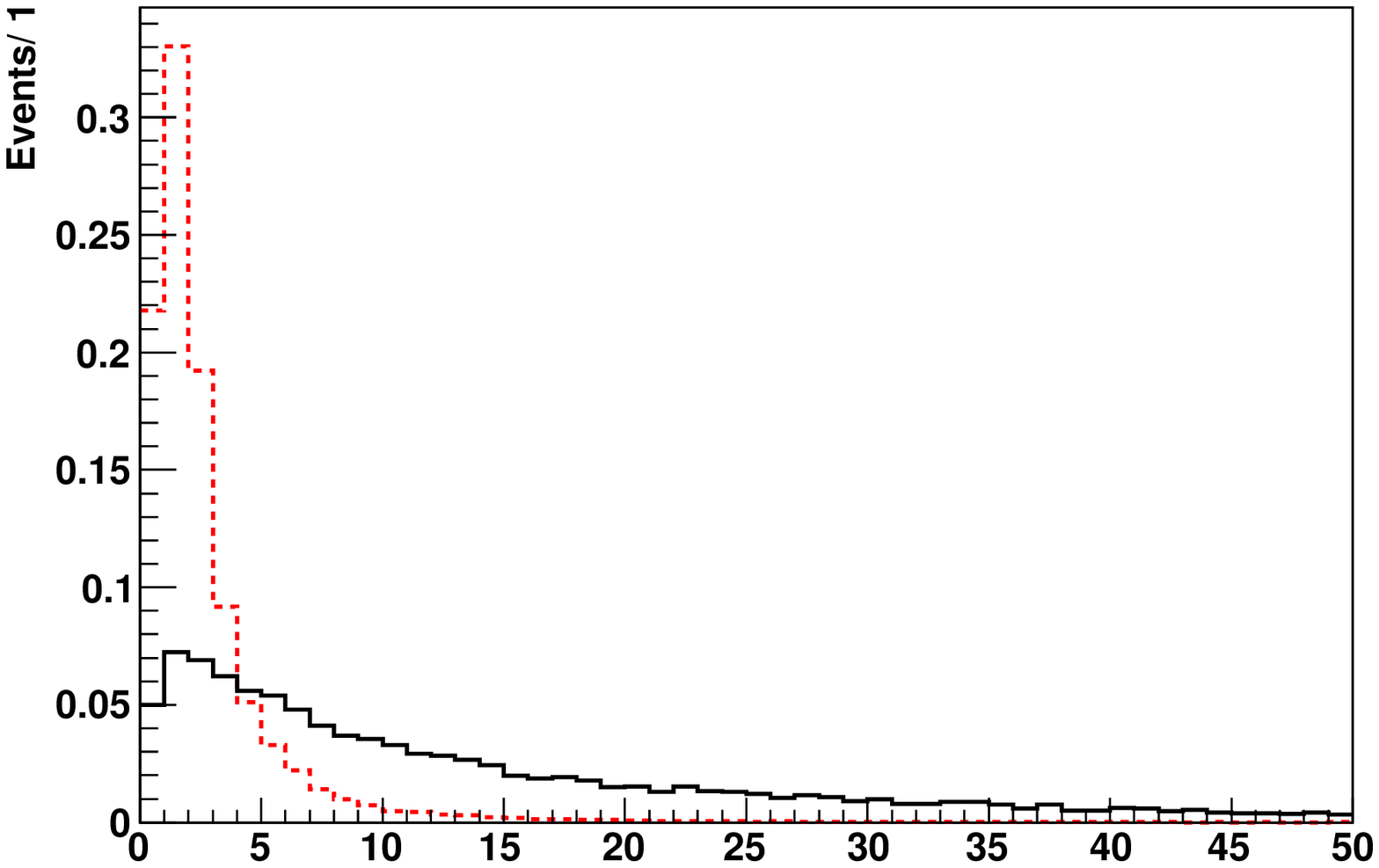}%
\includegraphics[width=0.43\textwidth]{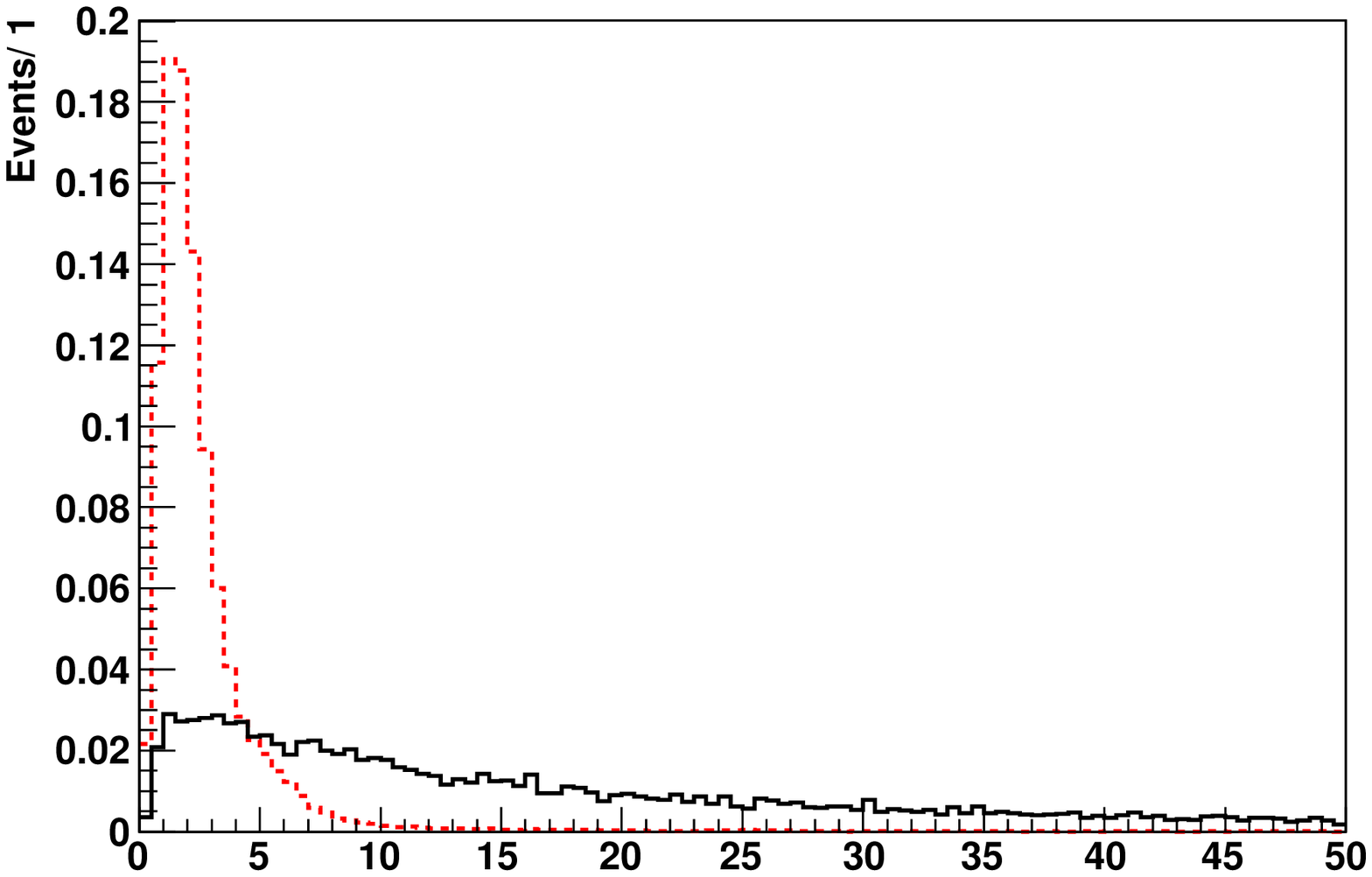}\break
\hbox to 0.43\textwidth{\hfil (a) $f_0$ minimum IPS   \hfil}%
\hbox to 0.43\textwidth{\hfil (b) $f_0$ flight significance
 \hfil}\break
\includegraphics[width=0.43\textwidth]{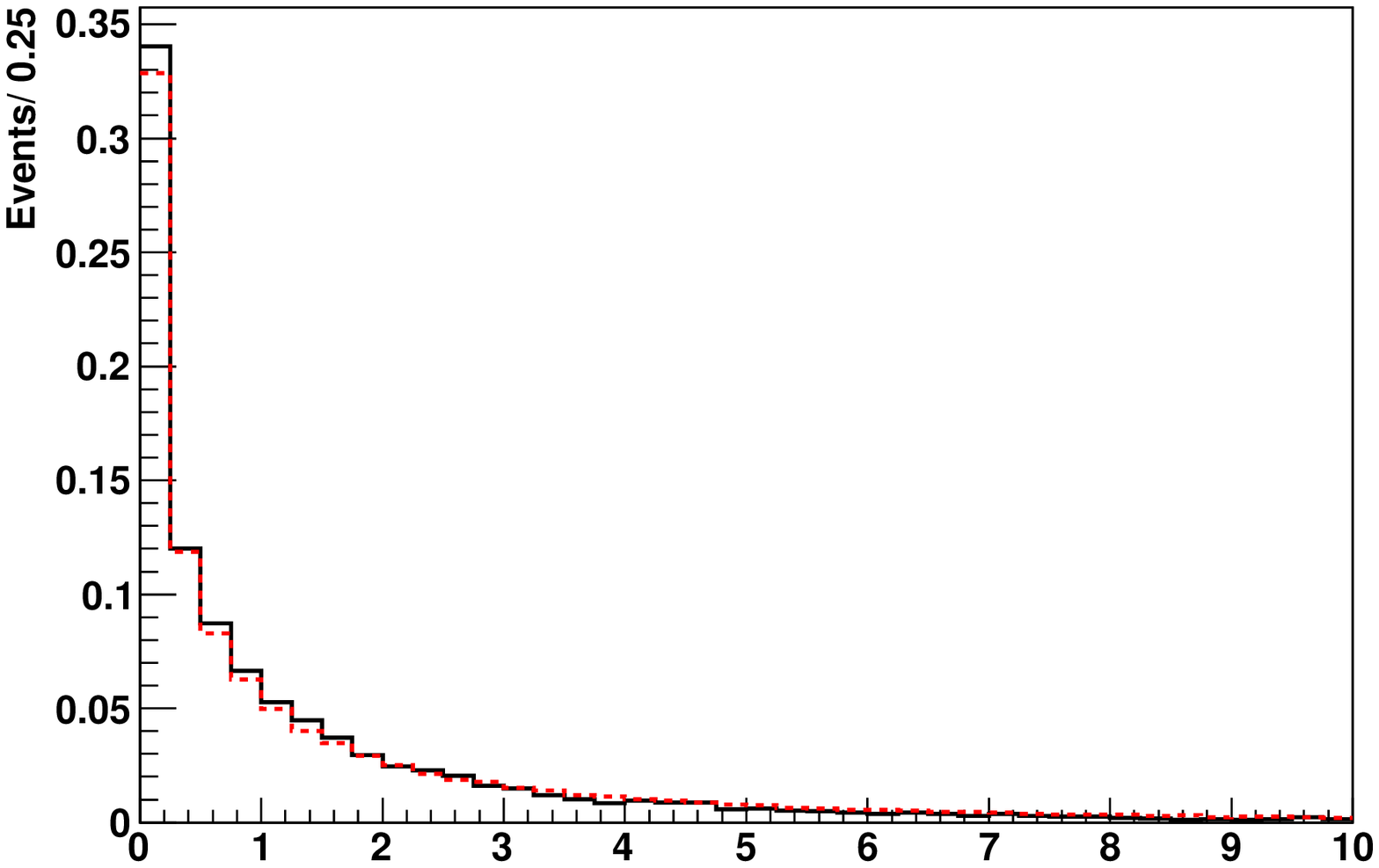}\break
\hbox to 0.43\textwidth{\hfil (c) $f_0$ vertex $\chi^2$ \hfil}\break
%\hbox to 0.43\textwidth{\hfil (d) $\Jpsi$ mass \hfil}\break

\caption{\label{dis-f0}The distributions of signal (black solid)
and $b\bar{b}$ background (red dashed) for $f_0(980)$ candidates.
(The number of events in the $B$ and $S$ distributions are equal.)}
\end{figure}

\subsection{\boldmath $\bs$ reconstruction}
A $\bs$ is reconstructed by combining $\Jpsi$ and $f_0(980)$ candidates. We require the four
tracks from $\Jpsi$ and $f_0(980)$ to be consistent with coming from one vertex as evaluated by examining the vertex fit $\chi^2$,
where the nDOF is 5. We also calculate the cosine of the angle between the $\bs$ candidates
reconstructed momentum and direction from the primary vertex to $\bs$ vertex ($\cos\theta_p$),
Fig.~\ref{dis-bs} shows the distributions of $\bs$ candidates from the signal and inclusive
$b\bar{b}$ MC after ``pre-selection."

\begin{figure}[htbp]
\center
\includegraphics[width=0.43\textwidth]{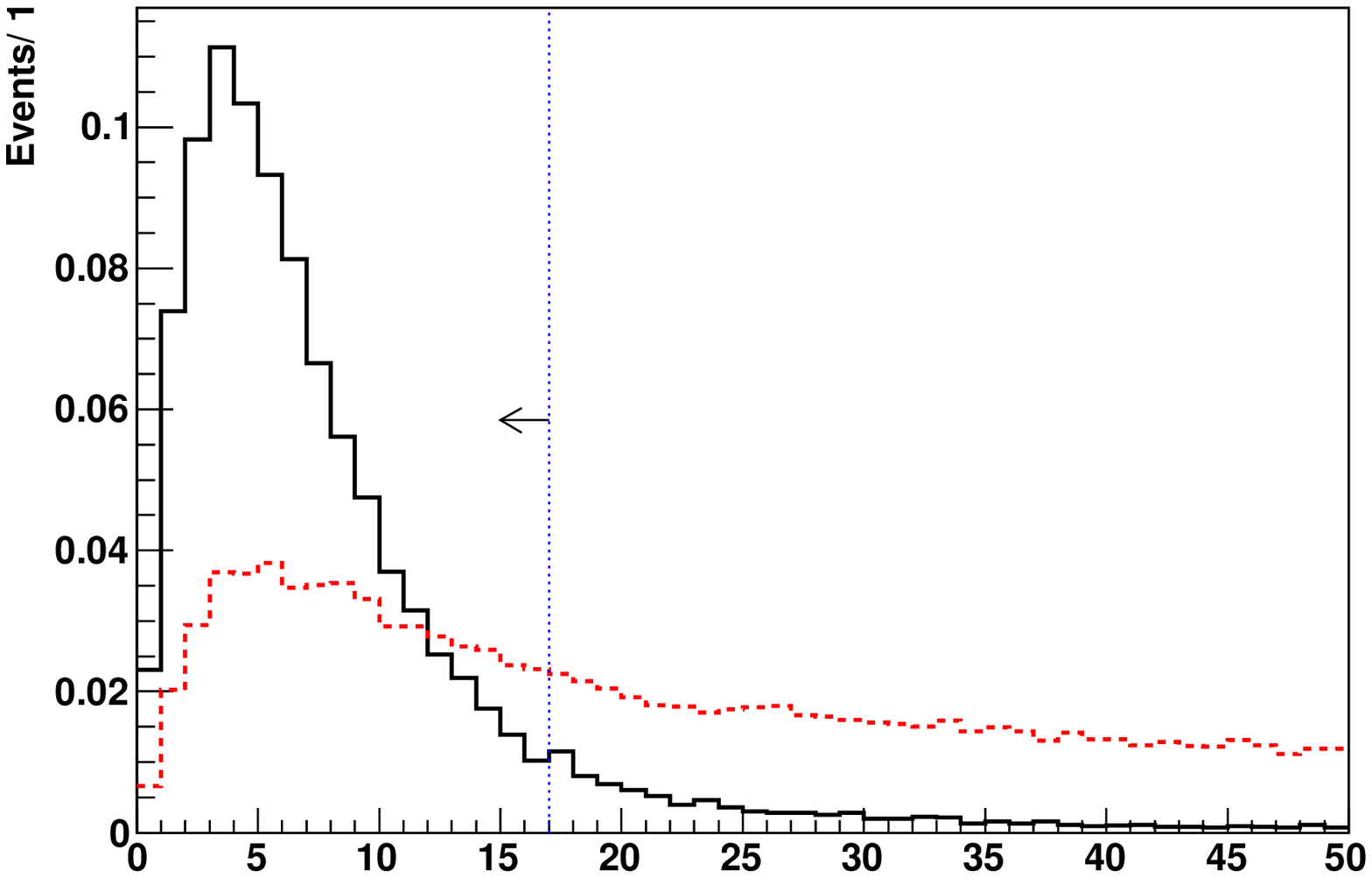}%
\includegraphics[width=0.43\textwidth]{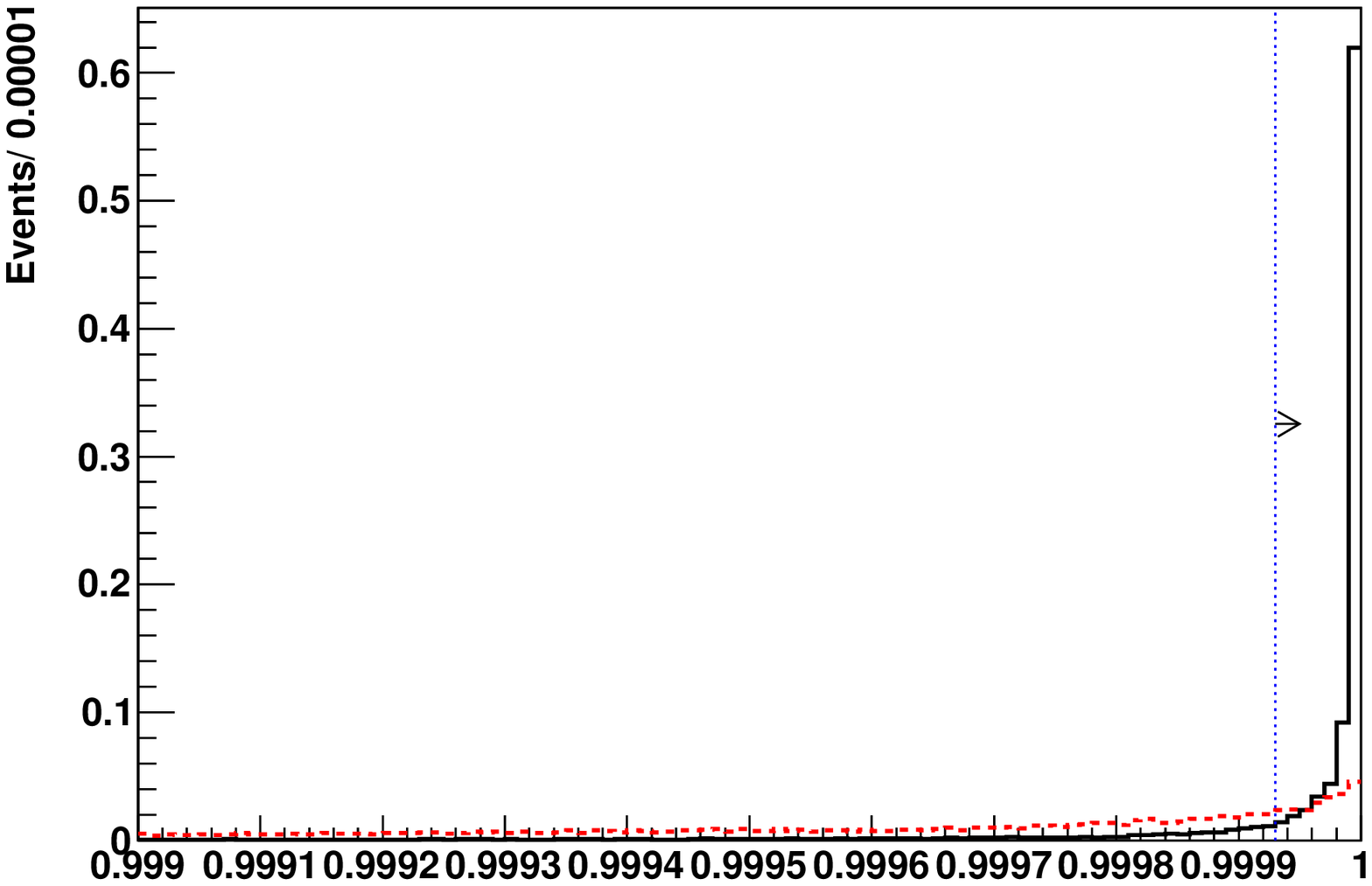}\break
\hbox to 0.43\textwidth{\hfil (a) vertex fit $\chi^2$   \hfil}%
\hbox to 0.43\textwidth{\hfil (b) $\cos\theta_p$ \hfil}\break
\includegraphics[width=0.43\textwidth]{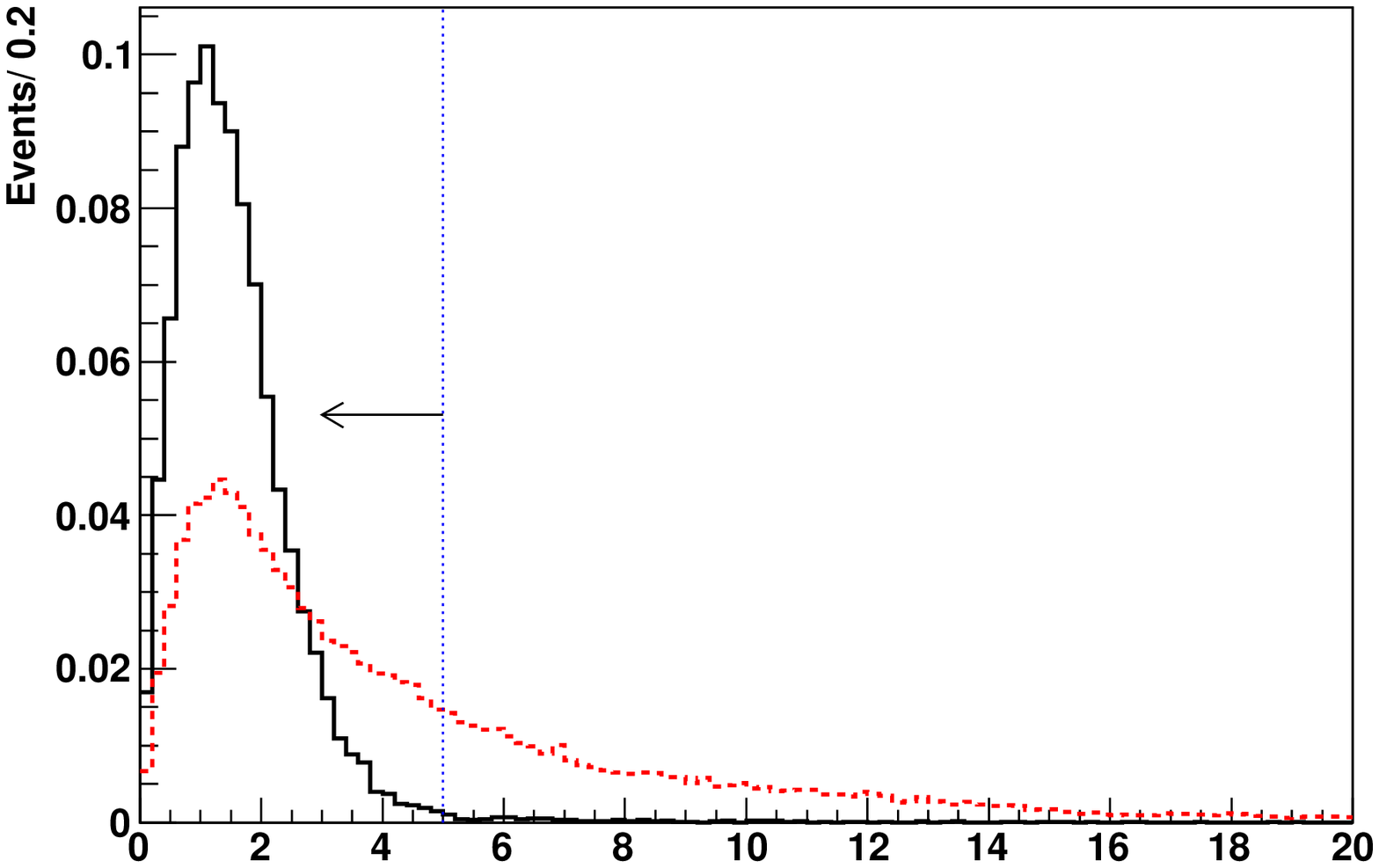}%
\includegraphics[width=0.43\textwidth]{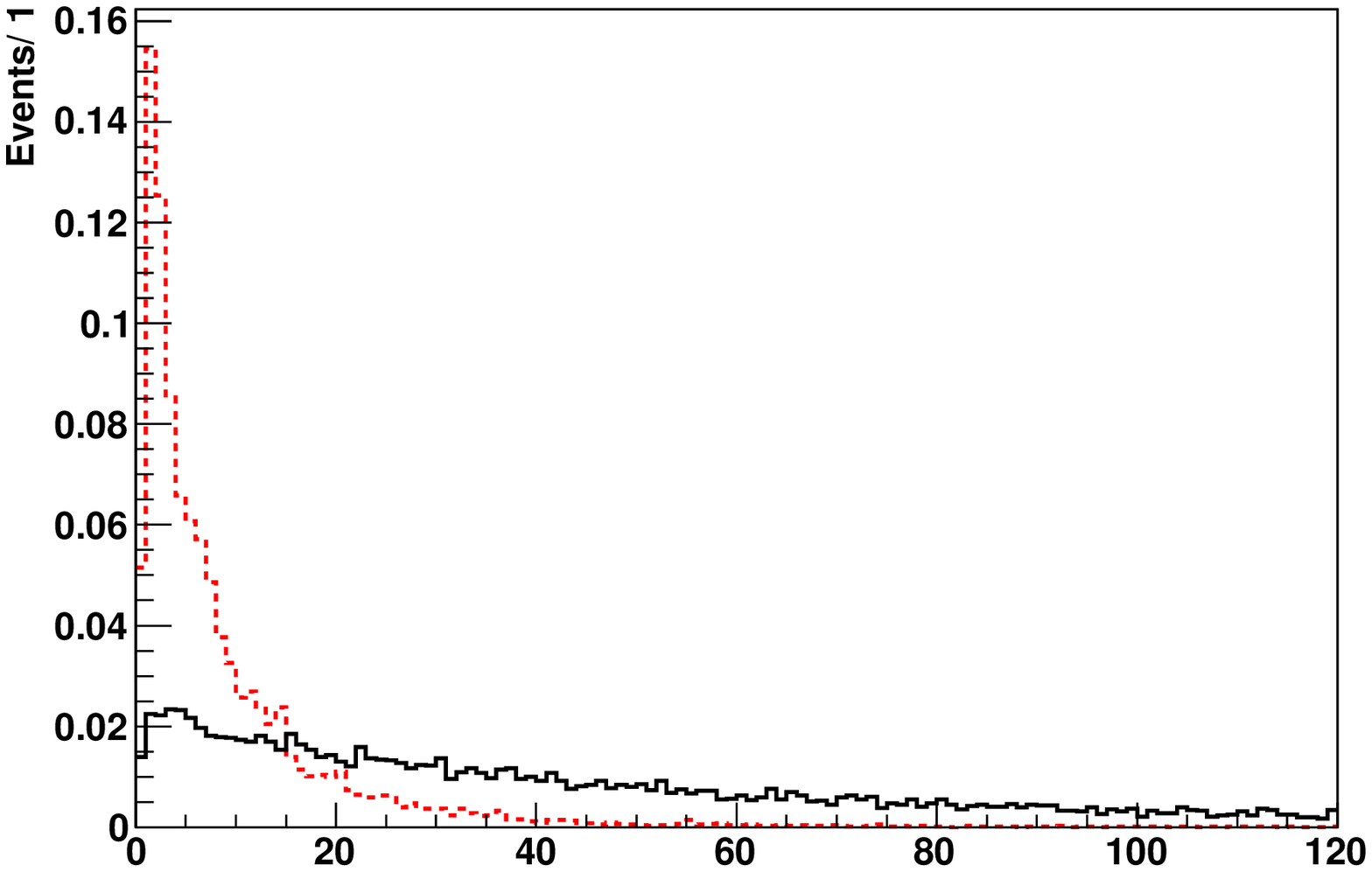}\break
\hbox to 0.43\textwidth{\hfil (c) IPS   \hfil}%
\hbox to 0.43\textwidth{\hfil (d) Flight significance \hfil}\break

\caption{\label{dis-bs}The distributions of the signal (black solid)
and $b\bar{b}$ background (red dashed) for $\bs$ candidates.
Final cuts on (a)-(c) are indicated by vertical blue dotted lines.}
\end{figure}

\subsection{Optimization Criteria}

To study how the $\phi_f$ sensitivity depends on the signal statistics
($S$) and the background to signal ratio ($B/S$), we use a fast stimulation
or ``Toy" Monte Carlo.
%The detail for the Toy MC generation is discussed later.

We generate and fit 400 ``experiments" as a function of the $B/S$ ratio, using $\phi_f$ set to
its predicted SM value of -0.0368. For each $B/S$ point we
fit the resulting $\phi_f$ distribution to a Gaussian.  (This is the same fitter as used in
section~\ref{sec:sens}.) The r.m.s. width ($\sigma$) of the
Gaussian is taken as the $\phi_f$ sensitivity. For our first point we fix $B/S$ at zero and
change only the amount of signal. Fig.~\ref{fom_s} shows the error of $\phi_f$
($\sigma_{\phi_f}$) as a function of signal yield. The curve follows the expected behavior:
$\sigma_{\phi_f}\propto 1/\sqrt{S}$. To understand the relation between $\sigma_{\phi_f}$ and
$B/S$, we fix $S$ and only change $B/S$. Fig.~\ref{fom_b} shows the the error of $\phi_f$
($\sigma_{\phi_f}$) as a function of $B/S$, where the experiment contains signal and a
long-lived background, with a lifetime fixed at 0.96 ps, as obtained from $B\to J/\psi X$ Monte
Carlo. We fit the data with a shape $\sigma_{\phi_f}\propto \sqrt{1+\alpha\times B/S}$. The value of $\alpha$ is
0.63$\pm$0.10, consistent with the ratio of background lifetime to $B_s$ lifetime, 0.64.
For an exponential proper time distribution the lifetime can be quickly estimated using
the mean of the distribution. For convenience we take
$\alpha$ as the ratio of mean values of the
proper-time distributions of background and signal.
(We have also found that the $\phi_f$ uncertainty due to the prompt $J/\psi$ background follows this formula.)

\begin{figure}[htbp]
\center
\includegraphics[width=0.55\textwidth]{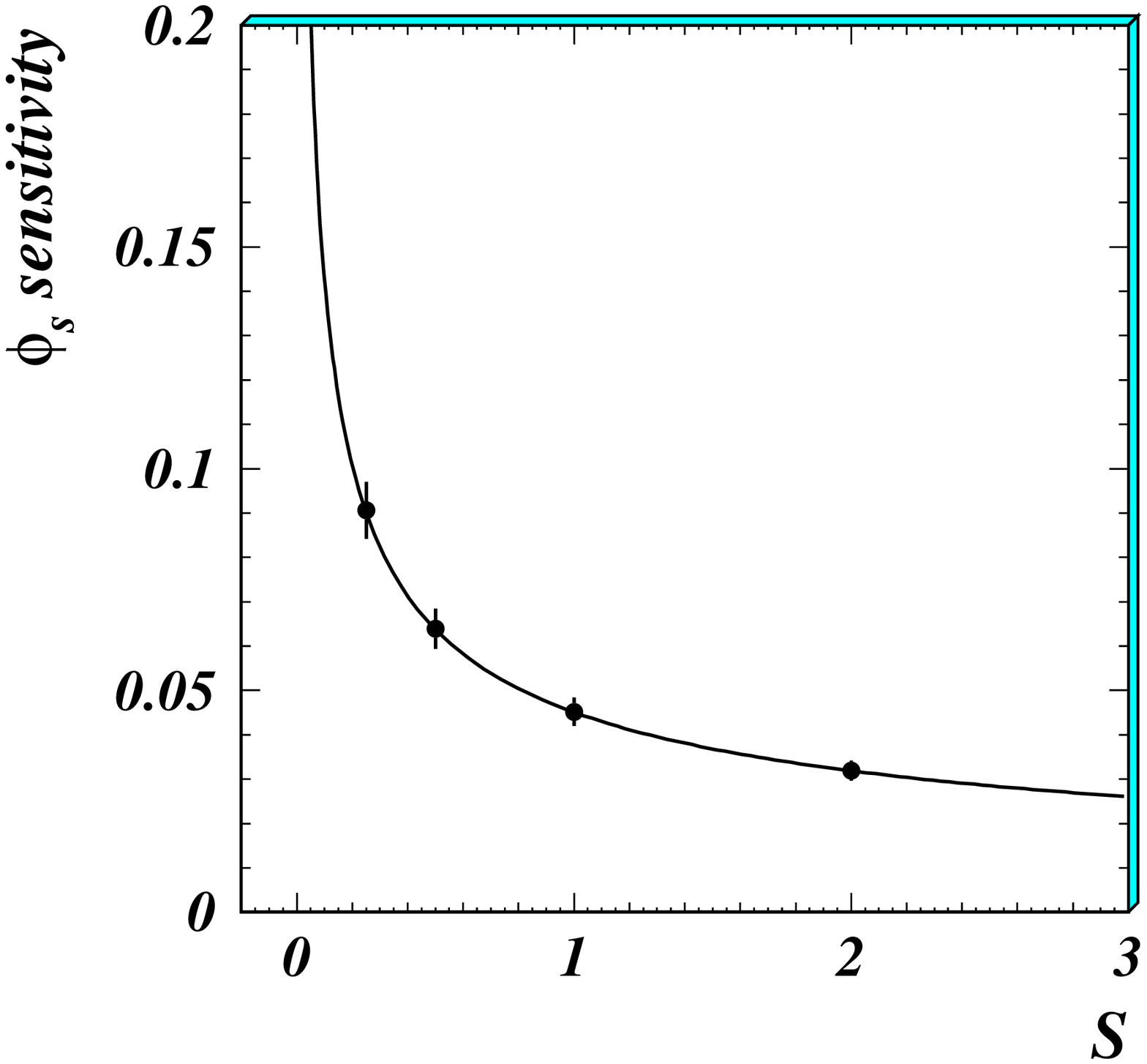}
\caption{\label{fom_s} The error of $\phi_f$ as a function of signal number $S$. ($S$ equal to one
corresponds to about 2 fb$^{-1}$ data.) The curve is the result of a fit to $\sigma_{\phi_f}\propto 1/\sqrt{S}$.}
\end{figure}

\begin{figure}[htbp]
\center
\includegraphics[width=0.55\textwidth]{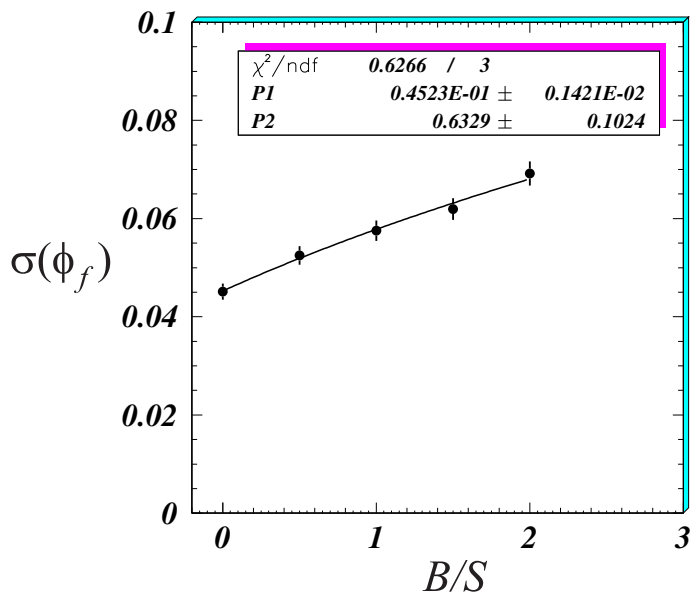}
\caption{\label{fom_b} The error of $\phi_f$ as a function of background to signal ratio $B/S$
for a fixed $S$. The $S$ corresponds to about 2 fb$^{-1}$ data. The curve shows a fit to
$\sigma_{\phi_f}=A\sqrt{1+\alpha\times B/S}$. Here $\alpha\equiv P2$ and $A\equiv P1$.}
\end{figure}

\subsection{Selection Optimization}
Our goal is to maximize the $\phi_f$ sensitivity.  Specifically, the selection cuts are chosen to maximum
$\frac{S}{\sqrt{S+\alpha\cdot B}}$ where $S$ ($B$) is the expected signal (background) number within
a $\pm 50$ MeV mass window around the $\bs$ nominal mass \cite{pdg08}, and $\alpha$ is the ratio of
the mean values of the proper-time distributions between the background and the signal. The cuts
are based on a set of variables that show a marked difference between signal versus background.
They include: (1) sum of absolute $p_{\rm T}$ of $f_0$ daughters (Fig. \ref{dis-pion} (a,b));
(2) impact parameter significance of pions (Fig. \ref{dis-pion} (c)); (3) track fit $\chi^2_{\rm
track}$/nDOF
 of pions (Fig. \ref{dis-pion} (d)); (4) vertex fit $\chi^2$ of $\mu^+\mu^-\pi^+\pi^-$
 (Fig. \ref{dis-bs} (a));
 and (5) the cosine of the
angle between the $\bs$ candidates reconstructed momentum and its flight direction ($\cos\theta_p$) (Fig.
\ref{dis-bs} (b)). (We have not explored optimizing the selection based on a neural
network or similar procedure as the tuning of such a method must be based on real data, and here
we are trying to understand the sensitivity to first order. If other variables
also have some discrimination power between signal and background, we will include them
in the future.)

 In what follows, we normalize all the yields to 2 fb$^{-1}$
data.
 We use the expected $b\bar{b}$
production cross-section $\sigma_{b\bar{b}}=500$ $\mu$b and $\sigma_{{\rm prompt}~ \Jpsi}=265.6$
$\mu$b as predicted by Pythia, and listed in Table~\ref{dc06}. Since the Monte Carlo used 698 $\mu$b as
the $b\bar{b}$ production cross-section, we scale the number of background events from $b\bar{b}$ to
the value corresponding to $\sigma_{b\bar{b}}=500$ $\mu$b. To have more statistics in the
background estimation, we enlarge the $f_0$ window from the nominal 90 MeV to 500 MeV, and the
$\bs$ window from 50 MeV to 300 MeV. If the background is distributed linearly in the larger
windows, we can scale the background number in the larger windows down by a factor of 33. The
factor $\alpha$ may change among cut variables, so we obtain $\alpha$ from the signal and background proper-time
distributions at each set of cut points. (We regard this as a minor point, and first analyses are likely to
use a fixed value of $\alpha$.)

We realize that the lower mass region contains a large $B^0 \to \Jpsi K^{*0}$ sample when the
kaon is misidentified as a pion (see Fig.~\ref{bd2jpsikst}a); there is also a small $B_{d}^0 \to \Jpsi \pi^+\pi^-$
background (see Fig.~\ref{expected_mass}), so we remove this background from the calculation
of $B$ as we will consider this background separately. The final cuts, called ``Selection," are shown in the third column
of Table~\ref{cuts}.

A selection cut of 900 MeV on $p_{\rm T}(\pi^+)+p_{\rm T}(\pi^-)$ serves to eliminate 42.4\% of
of the background, costing only 1.4\% of the signal efficiency.

\section{Signal efficiency and event yields}
Table \ref{eff} shows the efficiencies computed for $\bs \to \Jpsi
f_0$, where:
\begin{itemize}
\item $\epsilon_{\rm geo}$ is the efficiency of the acceptance cut at
generator level on the $\bs$ daughters between 10 to 400 mrad;

\item $\epsilon_{\rm det/geo}$ is the efficiency of that all $\bs$
decay products in the event are reconstructible (has enough MC hits
for long track reconstruction) ;

\item $\epsilon_{\rm rec/det}$ is the efficiency that the
reconstructible events are actually reconstructed.

\item $\epsilon_{\rm sel/rec}$ is the efficiency that the
reconstructed events are actually selected.

\item $\epsilon_{L0}$, $\epsilon_{HLT1}$, and $\epsilon_{HLT2}$ are respective
efficiencies for the L0, HLT1 and HLT2 trigger algorithms.
\end{itemize}

\begin{table}[htb]
\center
\caption{\label{eff}Efficiency for $\bs \to \Jpsi f_0(980)$. L0, HLT1 and HLT2 refer
to the three trigger steps.}
\begin{tabular}{ccccccc|c}
$\epsilon_{\rm geo}$ [\%]&$\epsilon_{\rm det/geo}$
[\%]&$\epsilon_{\rm rec/det}$ [\%]&$\epsilon_{\rm sel/rec}$ [\%]&
$\epsilon_{\rm L0}$ [\%]&$\epsilon_{\rm HLT1}$ [\%]&$\epsilon_{\rm HLT2}$ [\%]& $\epsilon_{\rm tot}$ [\%]\\\hline
16.4&38.7&71.4&45.4&94.0&95.0&95.0&1.75\\
\end{tabular}
\end{table}

The overall trigger efficiency is 85\%, for signal events with all decay products inside the detector.
This is a high efficiency, especially compared with fully hadronic events, which are typically at least
a factor of two lower.
The total efficiency including HLT2 is the product of the individual efficiencies listed above,
which is 1.75\%. The yield for a decay channel is computed as
\begin{equation}
Y = L_{\rm int} \times \sigma_{b\bar{b}} \times 2 \times f_{B_s} \times {\cal{B}}^{\rm vis}
\times \epsilon_{\rm tot},
\end{equation}
where $L_{\rm int}=2$ fb$^{-1}$ is the assumed integrated
luminosity, $\sigma_{b\bar{b}}=500$ $\mu$b is the expected $b\bar{b}$
production cross-section, $f_{B_s}
=(11.0\pm1.2)\%$ is the probability for a $b$-quark to form a $B_s$ meson, and
${\cal{B}}^{\rm vis}$ is the total visible branching fraction, the product of all the
individual branching fractions involved in the decay chain. (The factor of 2 arises because we can use either the $b$ or $\overline{b}$ quark.)
Assuming that  ${\cal B}(\bs \to \Jpsi f_0)\times {\cal B}(f_0\to \pi^+\pi^-)$ is quarter of ${\cal
B}(\bs \to \Jpsi \phi)\times {\cal B}(\phi\to K^+K^-)$, we have ${\cal{B}}^{\rm vis}(\bs \to
\Jpsi(\mu\mu) f_0(\pi\pi)) = 6.8\times10^{-6}$. After applying the efficiencies, we expect 26.1k signal events after
HLT2. Table~\ref{br} shows the branching and $b$-hadron production fractions assumed in the
calculation of the yields and background levels discussed below.

\begin{table}[htb]
\center
\caption{\label{br}Branching and $b$-hadron production fractions assumed in the calculation of
the yields and background levels.}
\begin{tabular}{|l|c|c|c|}\hline
Branching fraction& value& estimated from & Ref.\\\hline

 ${\cal{B}}^{\rm vis}(\bs \to
\Jpsi(\mu\mu) f_0(\pi\pi))$& $6.8\times10^{-6}$ &$\bs\to \Jpsi
\phi$&\cite{stone}\\\hline
 ${\cal{B}}^{\rm vis}(\bs \to
\Jpsi(\mu\mu) \eta^{\prime}(\rho \gamma))$& $10\times10^{-6}$&$\bd\to \Jpsi K^0$&\cite{btev}\\
\hline

${\cal{B}}^{\rm vis}(\bd \to \Jpsi(\mu\mu)
K^{*0}(K\pi))$&$(5.25\pm0.24)\times10^{-5}$&&PDG08
\cite{pdg08}\\\hline

${\cal{B}}^{\rm vis}(\bu \to \Jpsi(\mu\mu)
K^+)$&$(5.9\pm0.2)\times10^{-5}$&&PDG08 \cite{pdg08}\\\hline

 $f_u$& $(39.9\pm1.2)\%$& &PDG08
\cite{pdg08}\\\hline

$f_d$& $(39.9\pm1.2)\%$& &PDG08 \cite{pdg08}\\\hline

$f_s$& $(11.0\pm1.2)\%$ &&PDG08 \cite{pdg08}\\\hline
\end{tabular}

\end{table}

\section{Background Sources}

\subsection{\boldmath Prompt $\Jpsi$}
We use Monte Carlo corresponding to a 0.00125 fb$^{-1}$ data sample and find 69 candidates passing
the selection which are from prompt $\Jpsi$ events in the enlarged $f_0$ and $\bs$ mass regions, before applying the
trigger selections. From prompt $\Jpsi$ events we expect $B/S = (11\pm1)\%$, where the error reflects only the uncertainty due to Monte Carlo statistics.

\subsection{\boldmath Background from $b\overline{b}$}
Our Monte Carlo simulation used ${\cal{B}}(b\to \Jpsi X$)=1.46\% (see Table~\ref{dc06}), compared with the PDG value of
$(1.16\pm0.10)\%$ \cite{pdg08}. Using the Monte Carlo value, we expect $B/S = (26\pm2)\%$ from
$b\bar{b}$ background, where the error is only the statistical uncertainty from the Monte Carlo. This is reduced to $(20\pm 2)$\% using the PDG value. For the purpose of this note we will use the larger value, as that was what was used in previous studies of $J/\psi \phi$. We discuss specific background sources below.

\subsubsection{\boldmath $B_{u,d,s}\to\Jpsi X$}

Fig.~\ref{expected_mass} shows the $\mu^+\mu^-\pi^+\pi^-$
invariant mass distribution from $B_{u,d,s}\to\Jpsi X$.
\begin{figure}[htbp]
\center
\includegraphics[width=0.9\textwidth,height=0.4\textheight]{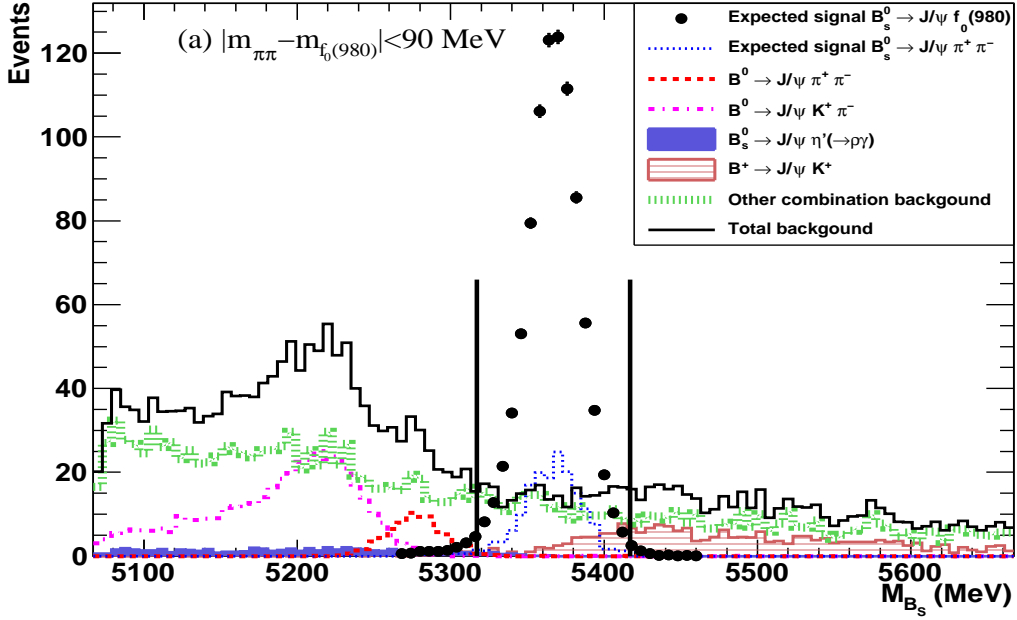}\break
\hbox to 0.9\textwidth{\hfil (a) $|m_{\pi\pi}-m_{f_0(980)}| < 90$
MeV \hfil}\break
\includegraphics[width=0.9\textwidth,height=0.4\textheight]{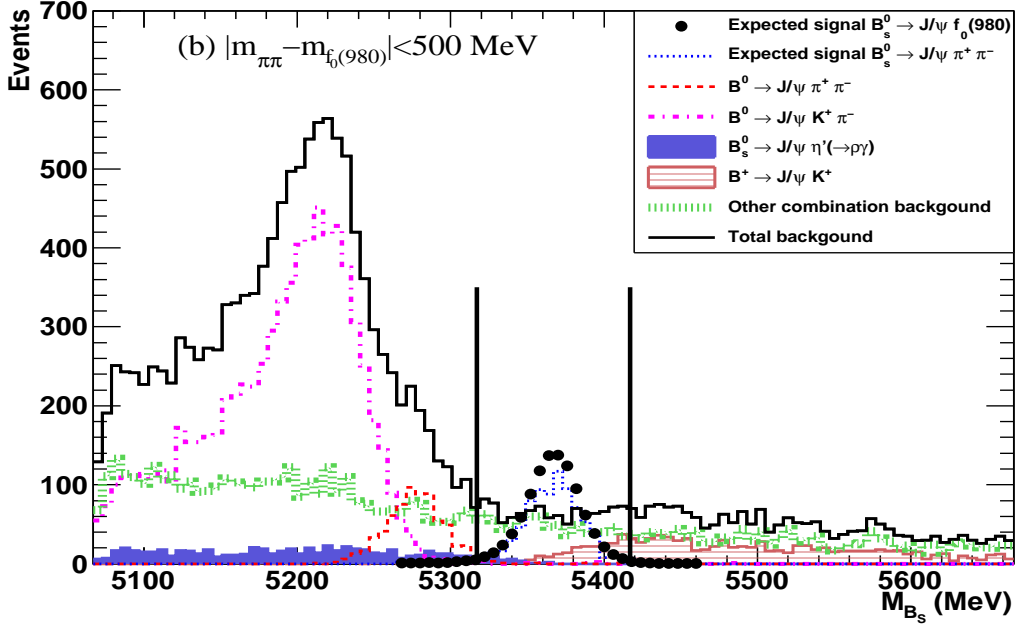}\break
\hbox to 0.9\textwidth{\hfil (b) $|m_{\pi\pi}-m_{f_0(980)}| < 500$
MeV \hfil}\break \caption{\label{expected_mass}
$\mu^+\mu^-\pi^+\pi^-$ invariant mass distributions from
$B_{u,d,s}\to\Jpsi X$ MC sample in (a) $\pm90$ MeV and (b) $\pm500$
MeV windows around the $f_0(980)$ nominal mass. The expected signal distributions
are superimposed. The non-resonant $\bs \to \Jpsi \pi^+ \pi^-$
branching fraction in this MC generation is set to $2\times10^{-4}$,
about 30\% of that of $\bs \to \Jpsi \phi, \phi\to K^+K^-$.}
\end{figure}
Fig.~\ref{f0mass_buds} shows the $m_{\pi\pi}$ distribution in the
$\bs$ mass signal region.
\begin{figure}[htbp]
\center
\includegraphics[width=0.5\textwidth]{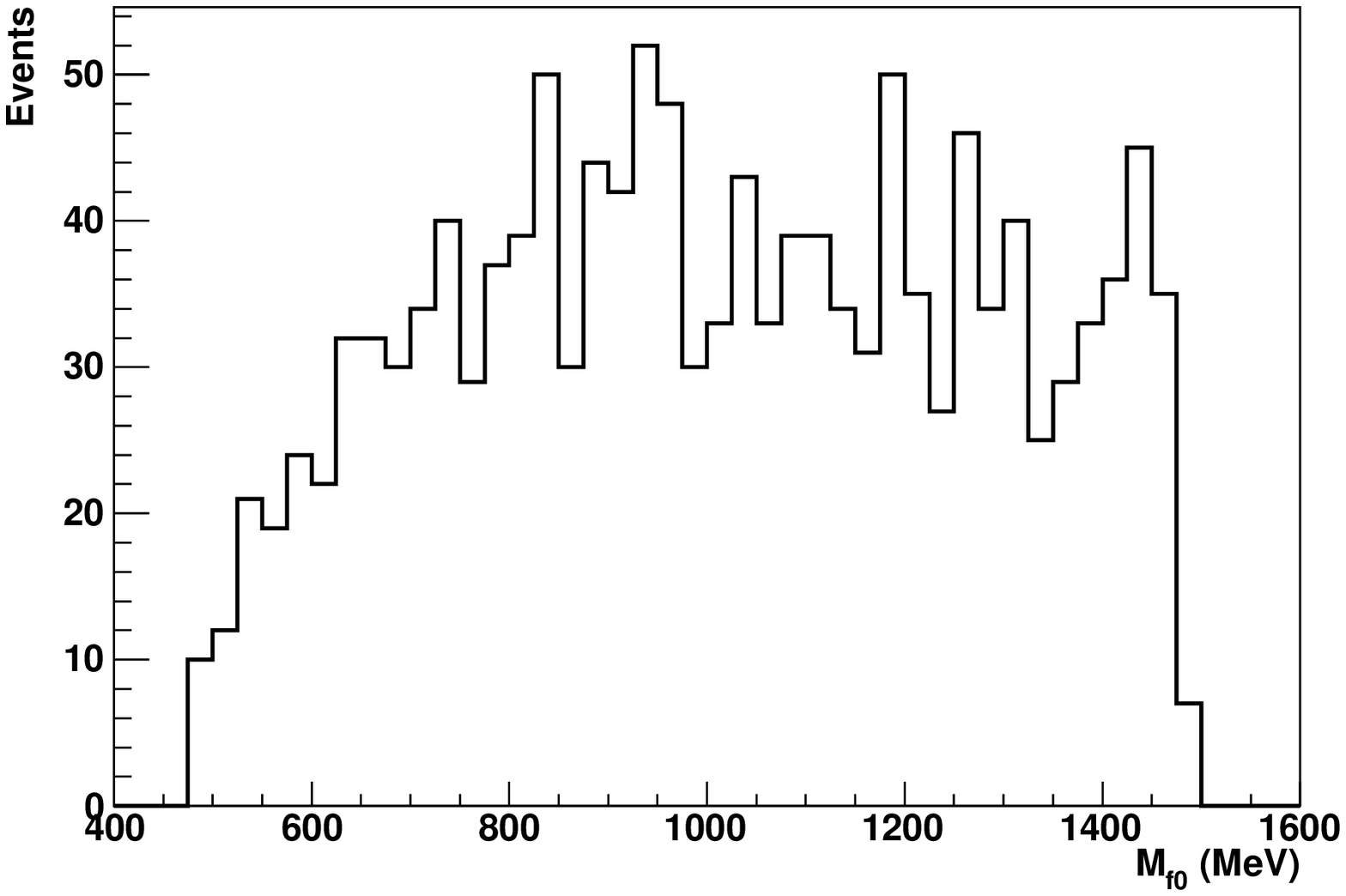}
\caption{\label{f0mass_buds}The $m_{\pi\pi}$ distribution in $\bs$
mass signal region ($|m_{\mu\mu\pi\pi}-m_{\bs}|<50$ MeV) from the sample
including all $B$ decays with $J/\psi$ candidates, except signal.}
\end{figure}

\subsubsection{\boldmath $\bd \to \Jpsi K^{*0}$}
When the kaon is misidentified as a pion, $\bd \to \Jpsi K^{*0}$ decay can fake a $B_s\to\Jpsi
\pi^+\pi^-$ event. Fig.~\ref{bd2jpsikst} shows the invariant mass distributions for $\bs$ and $f_0$ candidates
from $\bd \to \Jpsi K^{*0}$ decays. If the mass resolution does not deteriorate, we will see almost no events from this source; quantitatively, in 2 fb$^{-1}$ of data, we expect $<10$ events at 90\%
confidence level (C.L.) remaining in the $f_0$ and $\bs$ narrow signal
regions.
\begin{figure}[htbp]
\includegraphics[width=0.5\textwidth]{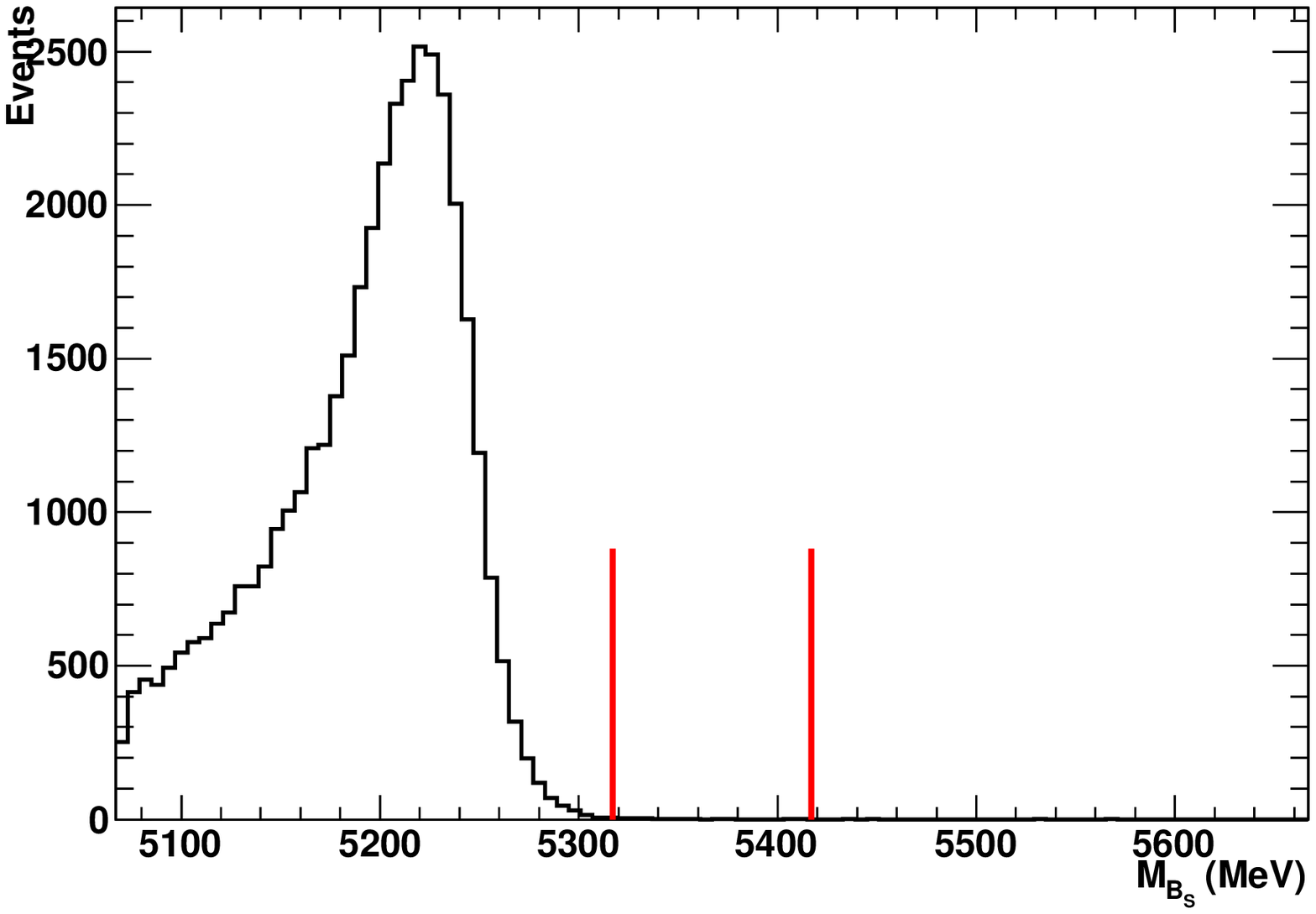}%
\includegraphics[width=0.5\textwidth]{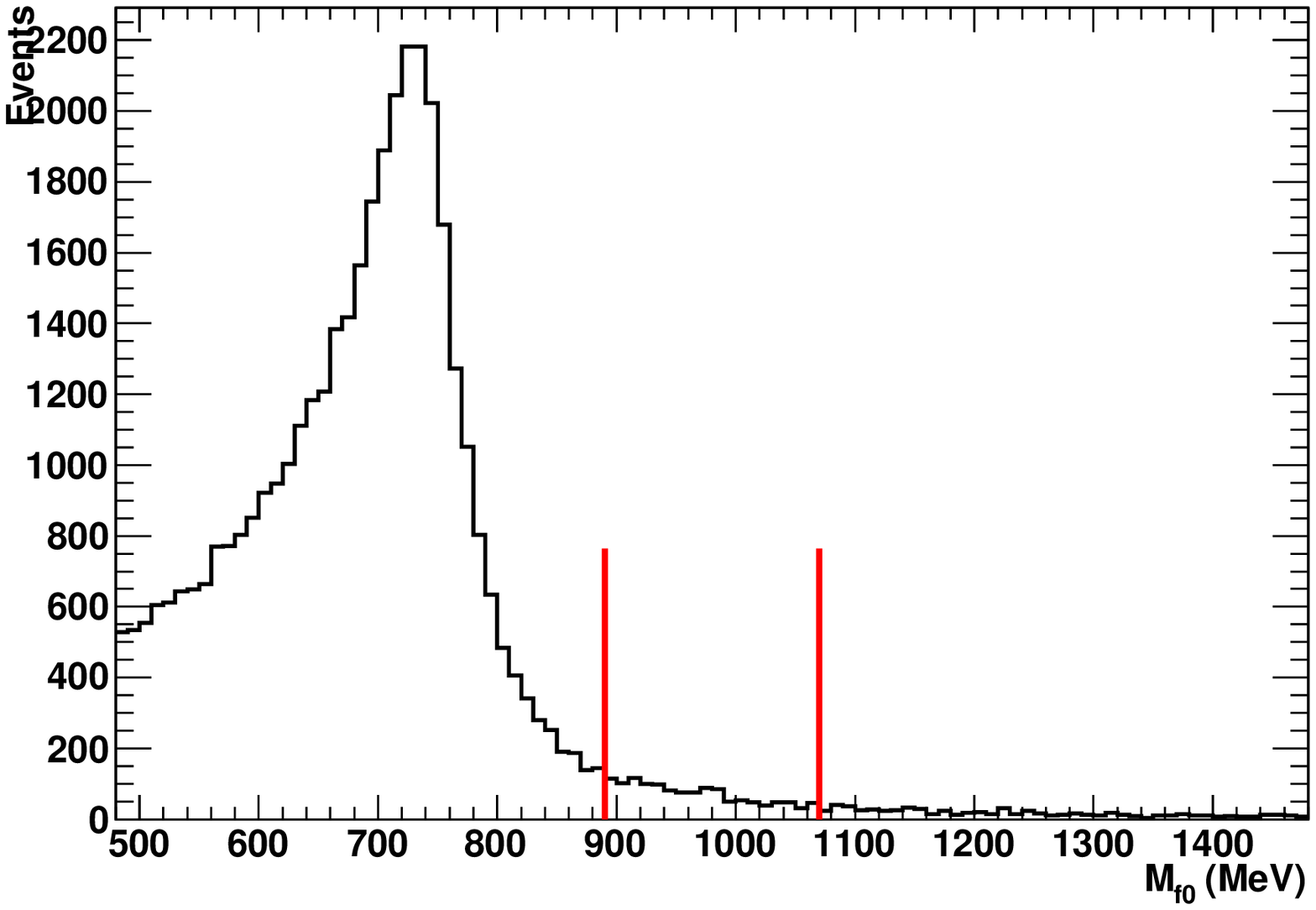}
\hbox to 0.5\textwidth{\hfil (a) $m_{\mu\mu\pi\pi}$  \hfil}%
\hbox to 0.5\textwidth{\hfil (b) $m_{\pi\pi}$\hfil}\break
\caption{\label{bd2jpsikst} Distributions from $\bd \to \Jpsi
K^{*0}$ decays when the kaon is misidentified as a pion. The vertical lines show the narrow
mass windows.}
\end{figure}

%$\bs \to \Jpsi \overline{K}^{*0}(K^-\pi^+)$ could have the similar
%distributions, and its branching fraction is expected to be
%4/3$\lambda^2$ (6.6\%) of that of $\bs \to \Jpsi \phi(K^-K^+)$.

\subsubsection{\boldmath $\bu \to \Jpsi K^+$}
The $\bu \to \Jpsi K^+$ decay can fake a $\Jpsi \pi^+\pi^-$ when the $K^+$ is misidentified as
pion and a random $\pi^-$ is combined. Fig.~ \ref{bd2jpsikst} shows the invariant mass
distributions for $\bs$ and $f_0$  from $\bu \to \Jpsi K^{+}$ decays. We expect
$(1.29\pm0.08)\times 10^{3}$ events contribute to the background before the trigger in 2 fb$^{-1}$
data.

\begin{figure}[htbp]
\includegraphics[width=0.5\textwidth]{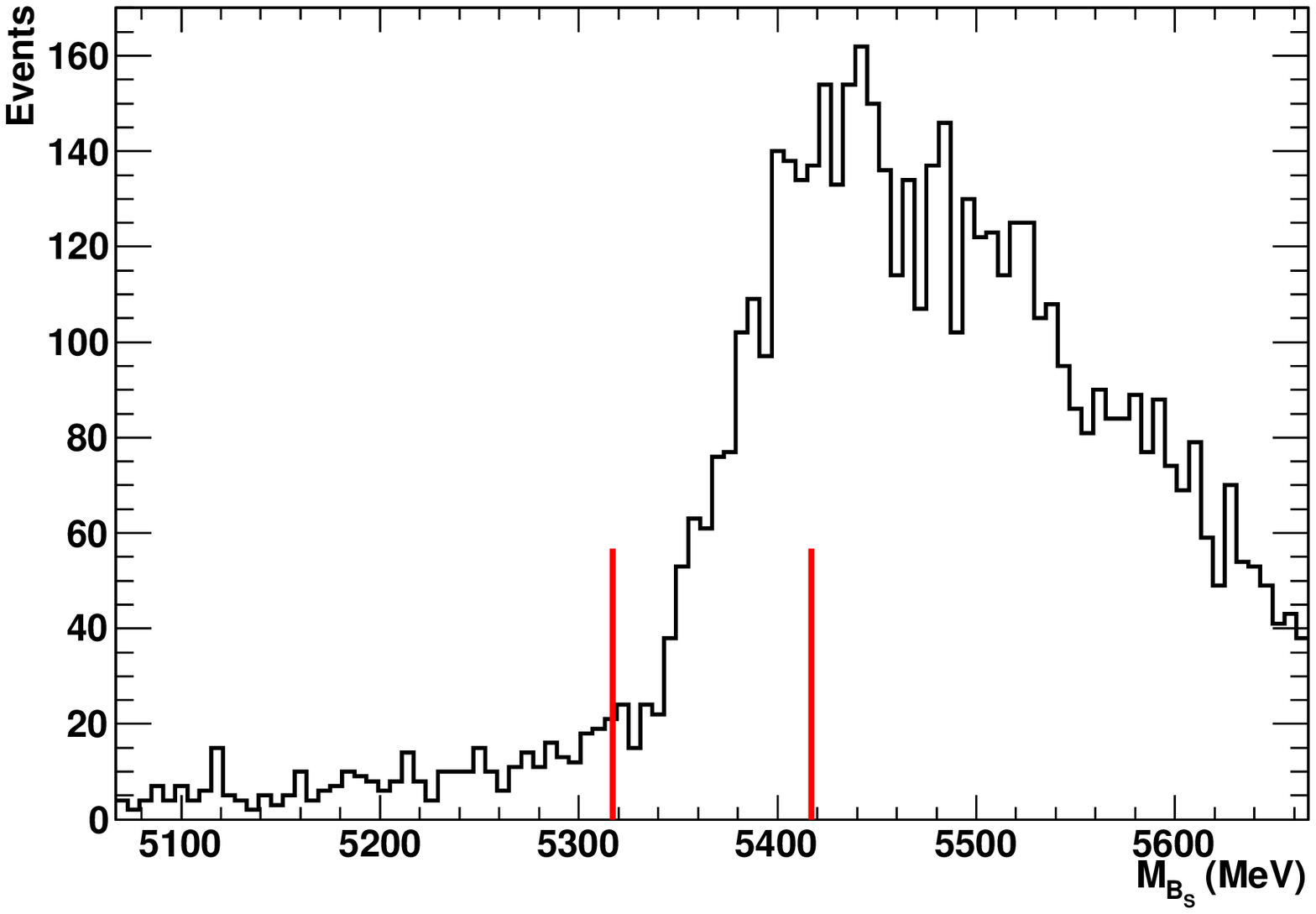}%
\includegraphics[width=0.5\textwidth]{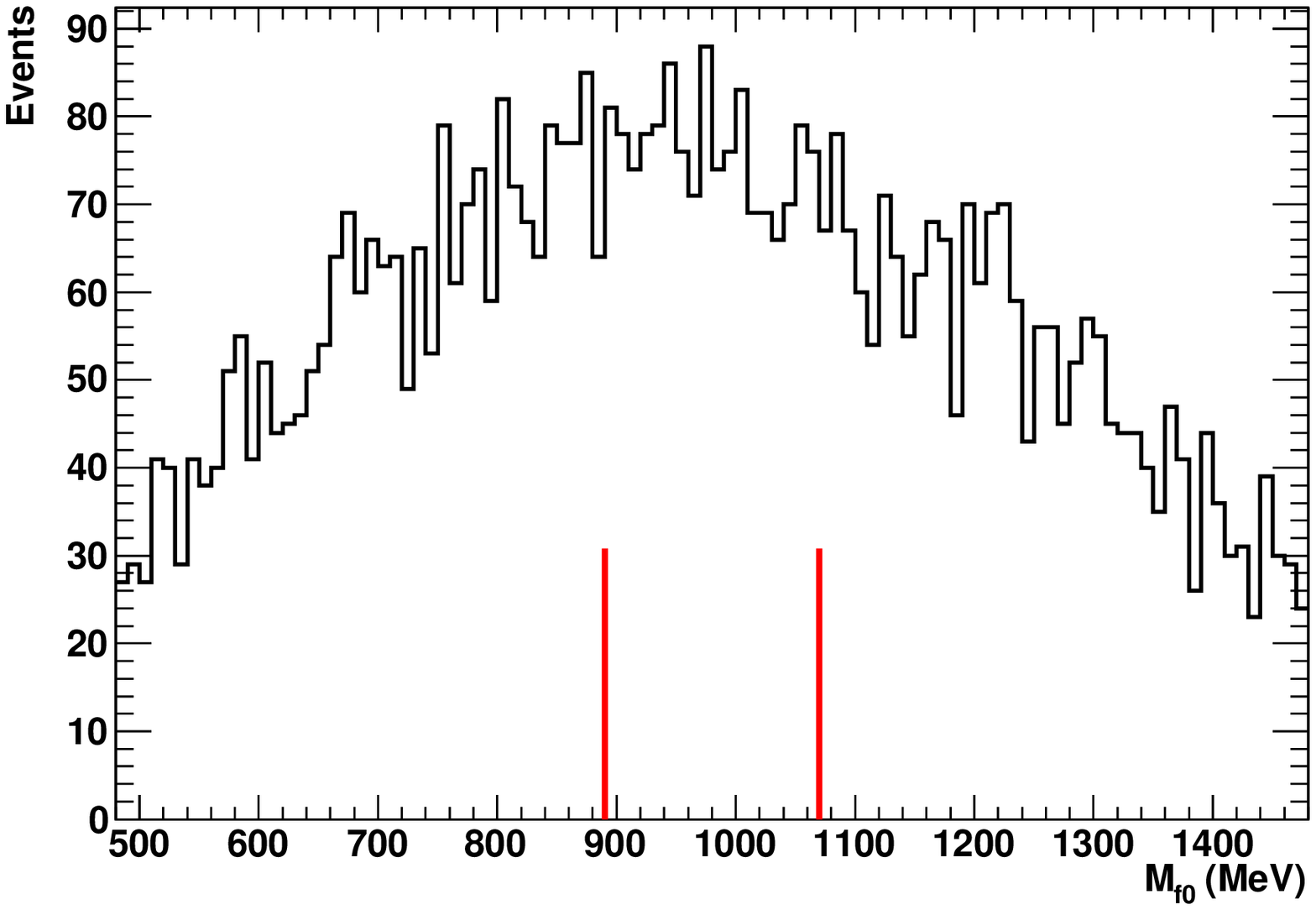}
\hbox to 0.5\textwidth{\hfil (a) $m_{\mu\mu\pi\pi}$  \hfil}%
\hbox to 0.5\textwidth{\hfil (b) $m_{\pi\pi}$\hfil}\break
\caption{\label{bu2jpsik} Distributions from $\bu \to \Jpsi K^+$
decays where the $K^+$ is misidentified as pion and a random $\pi^-$ is
combined. The vertical lines show the narrow
mass windows.}
\end{figure}

\subsubsection{\boldmath $\bs \to \Jpsi \eta^{\prime}$}
We have studied $\eta^{\prime} \to \eta \pi^+\pi^-$ and $\eta^{\prime} \to \rho \gamma$ and
found only the latter could contribute as background. It is a dangerous background, because
$\bs \to \Jpsi \eta^{\prime}$ has opposite CP to the signal. Fig.~ \ref{bs2jpsietap}
shows the invariant mass distributions for $\bs$ and $f_0$  from $\bs \to \Jpsi \eta^{\prime}$,
$\eta^{\prime} \to \rho \gamma$ decays. Assuming ${\cal{B}}(\bs \to \Jpsi \eta^{\prime})$ =
$\frac{2}{3} $ ${\cal{B}}(B^0 \to \Jpsi K^0)$ \cite{btev}, we expect $(0.75\pm0.07)\times10^{3}$ events
contribute to the background, before the trigger, in 2 fb$^{-1}$ data. Most of these, however,
are not in the narrow signal mass ranges.

We also intend to reconstruct these events when we can find the photon from the $\eta'\to \rho\gamma$ decay to use in a separate measurement of $\phi_f$ (see Sec.~\ref{subsec:etap}).  We have not vetoed these events in our $J/\psi f_0$ selection, in this study.

\begin{figure}[htbp]
\includegraphics[width=0.5\textwidth]{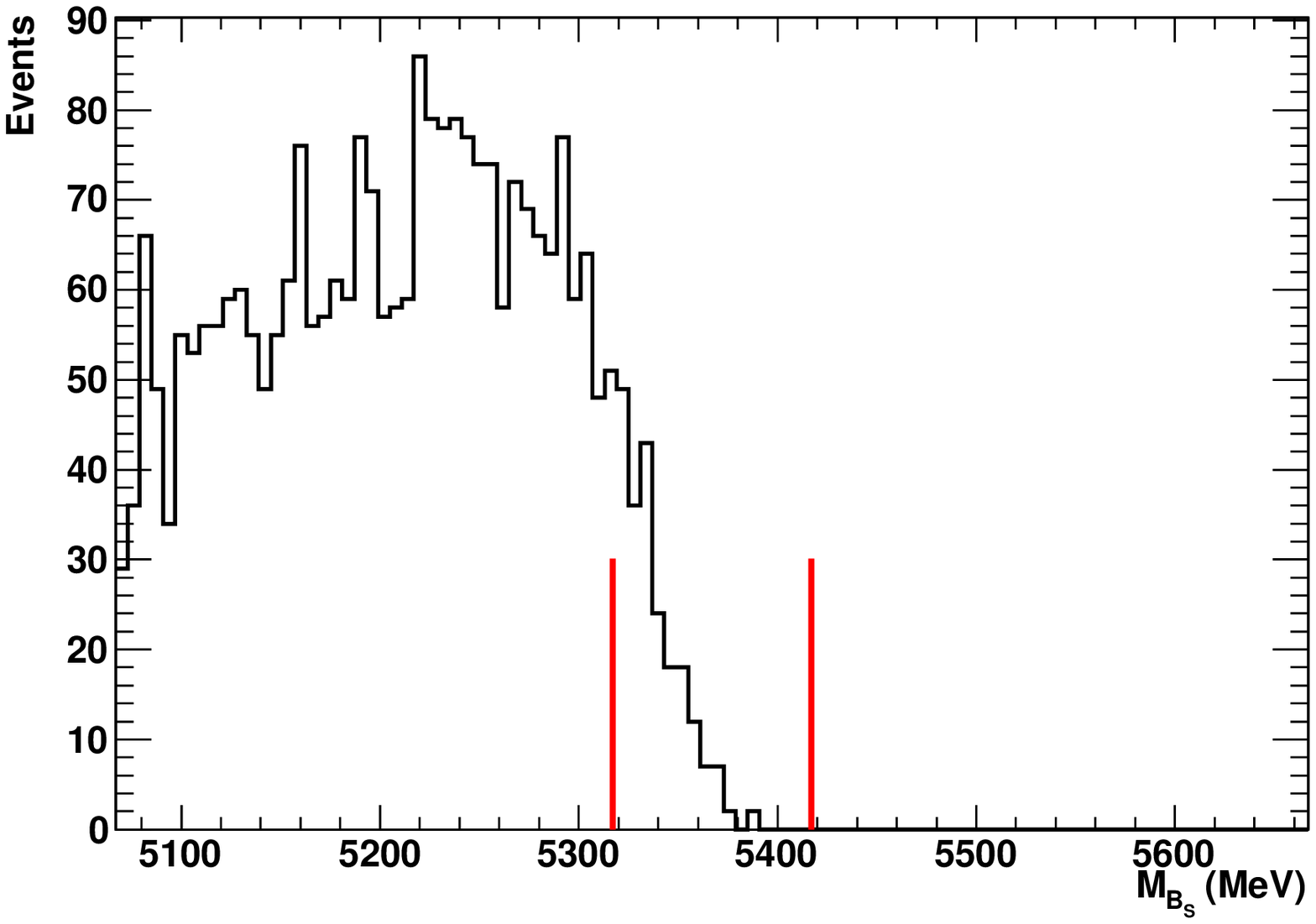}%
\includegraphics[width=0.5\textwidth]{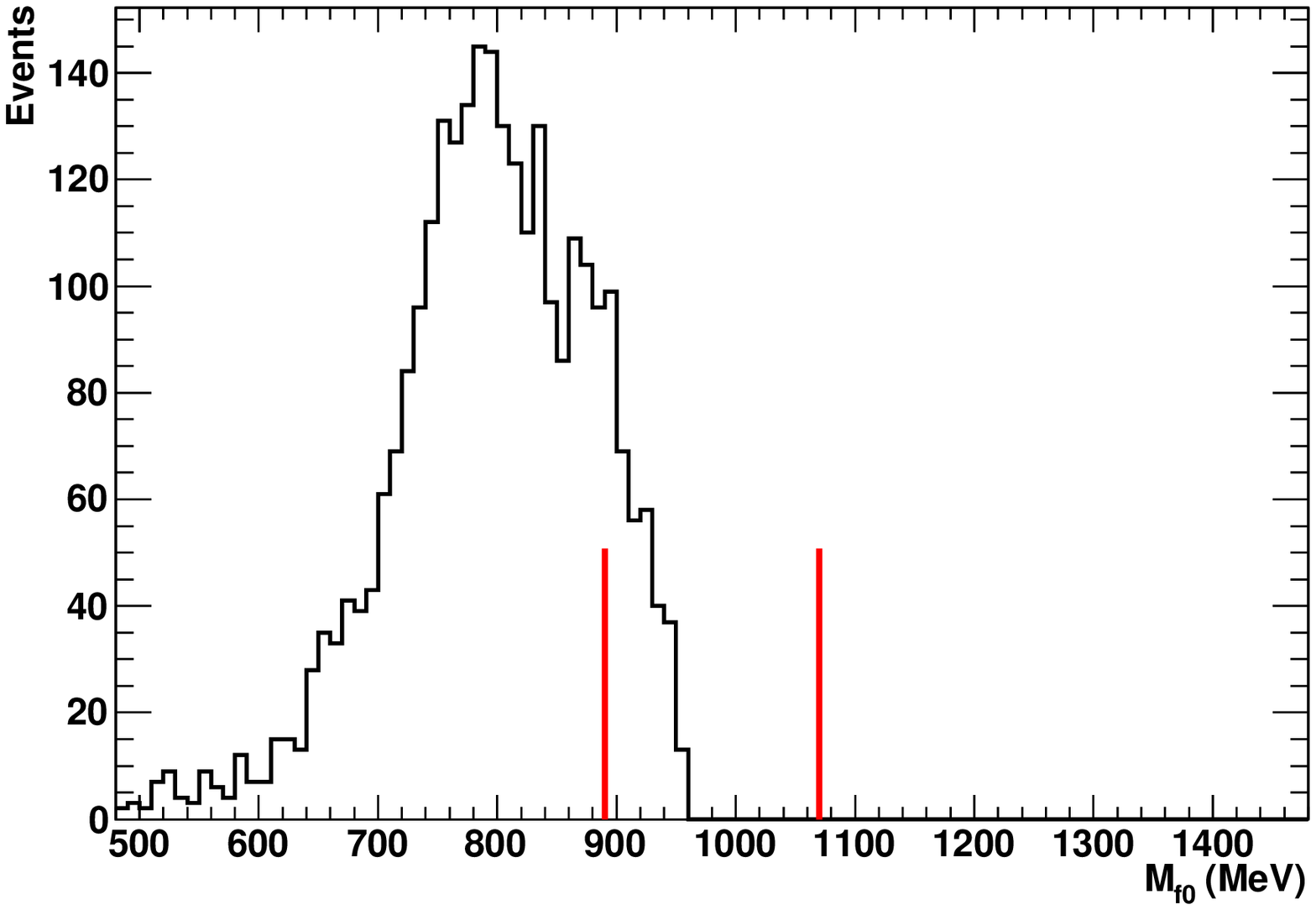}
\hbox to 0.5\textwidth{\hfil (a) $m_{\mu\mu\pi\pi}$  \hfil}%
\hbox to 0.5\textwidth{\hfil (b) $m_{\pi\pi}$\hfil}\break
\caption{\label{bs2jpsietap} Distributions from $\bs \to \Jpsi
\eta^{\prime}$, $\eta^{\prime} \to \rho \gamma$  decays. The vertical lines show the narrow mass windows.}
\end{figure}

\subsection{Background Summary}
\begin{table}[htb]
\center
\caption{\label{bg}Summary of background sources. The numbers quoted are
calculated in the narrow mass windows $|m_{\pi\pi}-m_{f_0(980)}|<90$ MeV and
$|m_{\mu\mu\pi\pi}-m_{\bs}|<50$ MeV. We assume the
trigger efficiencies are the same for the signal and backgrounds. The $b\bar{b}$ background does not include the specific
$\bs$, $\bu$, $\bd$ channels listed here.}
\begin{tabular}{c|c|c}\hline
Sources & Yield from 2 fb$^{-1}$ before trigger ($\times 10^3$)& $B/S$\\\hline\hline
 signal &
$29.3$\\\hline\hline background& & \\\hline
prompt $\Jpsi$ & $3.3\pm0.4$&$(11\pm1)\%$\\
$b\bar{b}$&$6.6\pm0.5$&$(21\pm2)\%$\\
$\bs \to \Jpsi \eta^{\prime}$&$0.75\pm0.07$&$(2.4\pm 0.2)$\%\\
$\bu \to \Jpsi K^+$&$1.29\pm0.08$& $(4.2\pm 0.3)$\%\\
$\bd \to \Jpsi K^{*0}$&$<10$ $@$90\%C.L.&\\\hline
Total &  $11.3\pm 0.6$ & ($39\pm3$)\%\\
\hline
\end{tabular}

\end{table}

Table~\ref{bg} shows the sources of the background and expected yields from 2 fb$^{-1}$ data.
The branching fractions and $b$-hadron production fractions used for the calculation are shown in Table
\ref{br}. The largest background arises from a $\Jpsi$ combined with random tracks identified as
pions, other than $\bd \to \Jpsi K^{*0}$ and $\bs \to \Jpsi \eta^{\prime}$. These backgrounds
(mass shape and proper time distribution) can be estimated by reconstructing $\Jpsi$ with two
same sign-charged pions; we cannot use the $\bs$ mass sidebands due to reflections from specific
final states. We show in Fig.~\ref{mass_wch2} the $J/\psi\pi\pi$  mass distribution for like sign and opposite sign pion pairs.
In the $\bs$ signal region, and even above, the distributions are in agreement.

\begin{figure}[htbp]
\center
\includegraphics[width=0.5\textwidth]{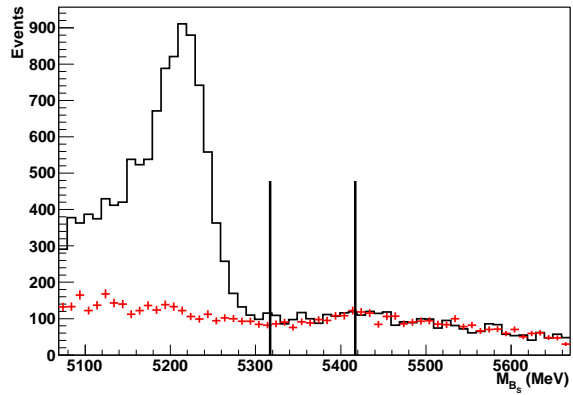}
\caption{\label{mass_wch2} Invariant mass distributions from from $\bs \to \Jpsi\pi\pi$  candidates. The Monte Carlo sample consists of any $B$ decay into a final state containing a $J/\psi$ where the modes $\bs\to J/\psi f_0$,  $\bs\to J/\psi \eta'$ and $\bd\to J/\psi \pi^+\pi^-$ have been explicitly removed. The histogram is for $\pi^+\pi^-$ pairs, while the crosses represent $\pi^{\pm}\pi^{\pm}$ pairs.  The vertical lines define the $\bs$ signal window .}
\end{figure}

Furthermore, as a check, we processed a 5.6 million mini-bias event sample that satisfied the L0 trigger.
Only 2 events passed the off-line selection, and they both contain real  $J/\psi$ decays.

\section{Properties of selected signal events}
\subsection{\boldmath $\bs$ mass resolution}
Fig.~\ref{mass} shows the reconstructed $\bs$ mass distribution for the signal. Performing a
double-Gaussian fit gives $\bar\sigma(M_{\bs})=19$ MeV, where the average width
$\bar\sigma=\sqrt{(1-f_2)\sigma_1^2+f_2\sigma_2^2}$, $\sigma_1$ and $\sigma_2$ are the width of
the core and second Gaussian, and $f_2$ is the fraction of the second Gaussian \cite{constraint}.
\begin{figure}[htbp]
\center
\includegraphics[width=0.55\textwidth]{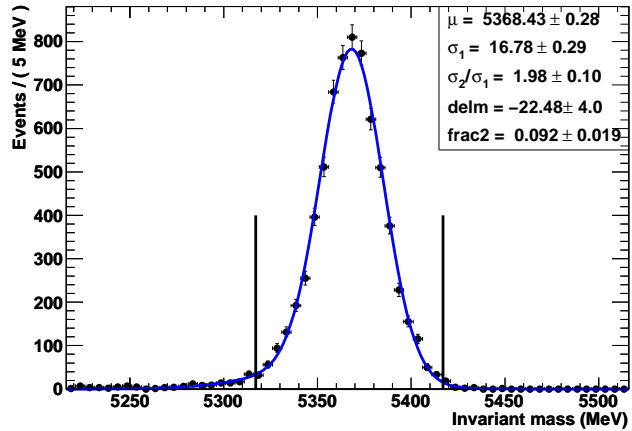}
\caption{\label{mass} Reconstructed $\bs$ mass for the signal. The
HLT1 trigger has been applied.}
\end{figure}

\subsection{Proper time}
The $\bs$ proper time is defined here as:
\begin{equation}
t^{\rm rec} = m \cdot \frac{\vec{d}\cdot\vec{p}}{|p|^2},
\end{equation}
where $m$ is the reconstructed invariant mass, $\vec{p}$ the
momentum and $\vec{d}$ the distance of flight vector of the
candidate $\bs$ from the primary to the secondary vertices. If more
than one primary vertex is found, the one that corresponds to the
smallest IP of the $\bs$ candidate is chosen.

 Fig.~\ref{dt} shows the distribution of the difference of the
reconstructed ($t^{\rm rec}$) and MC true ($t^{\rm MC}$) proper time. The average proper time resolution
obtained from a double-Gaussian fit is 34 fs.

\begin{figure}[htbp]
\center
\includegraphics[width=0.55\textwidth]{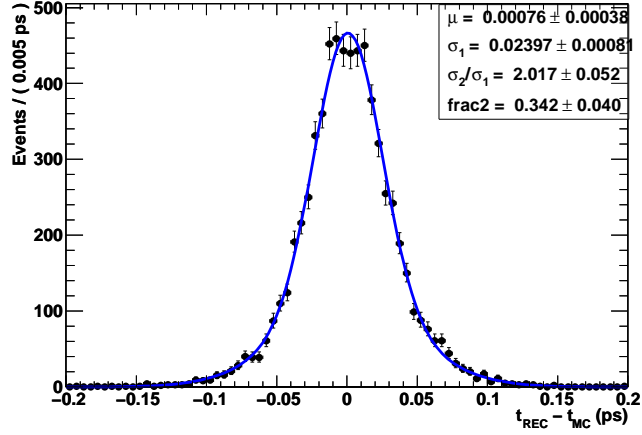}
\caption{\label{dt}The distribution of the difference between the
reconstructed ($t^{\rm rec}$) and MC true ($t^{\rm MC}$) proper
time. The HLT1 trigger has been applied.}
\end{figure}

Fig.~\ref{pull} shows the proper time error estimate and the
proper time pull $\frac{t^{\rm rec}-t^{\rm MC}}{\sigma_t}$.  The
estimate of the error on the reconstructed $\bs$ lifetime has a mean
value of 27 fs and a most probable value of 25 fs.
\begin{figure}[htbp]
\center
\includegraphics[width=0.45\textwidth]{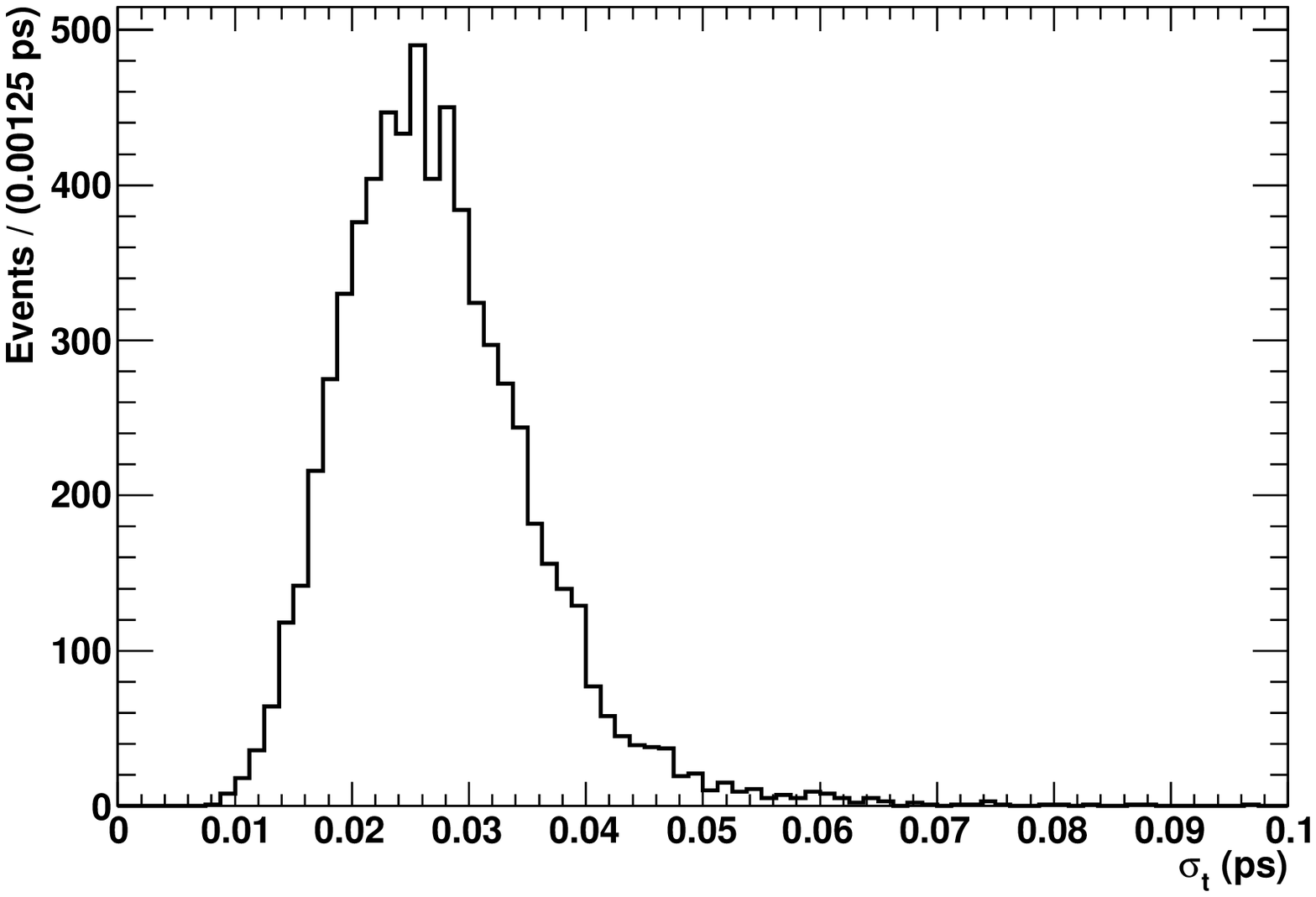}%
\includegraphics[width=0.45\textwidth]{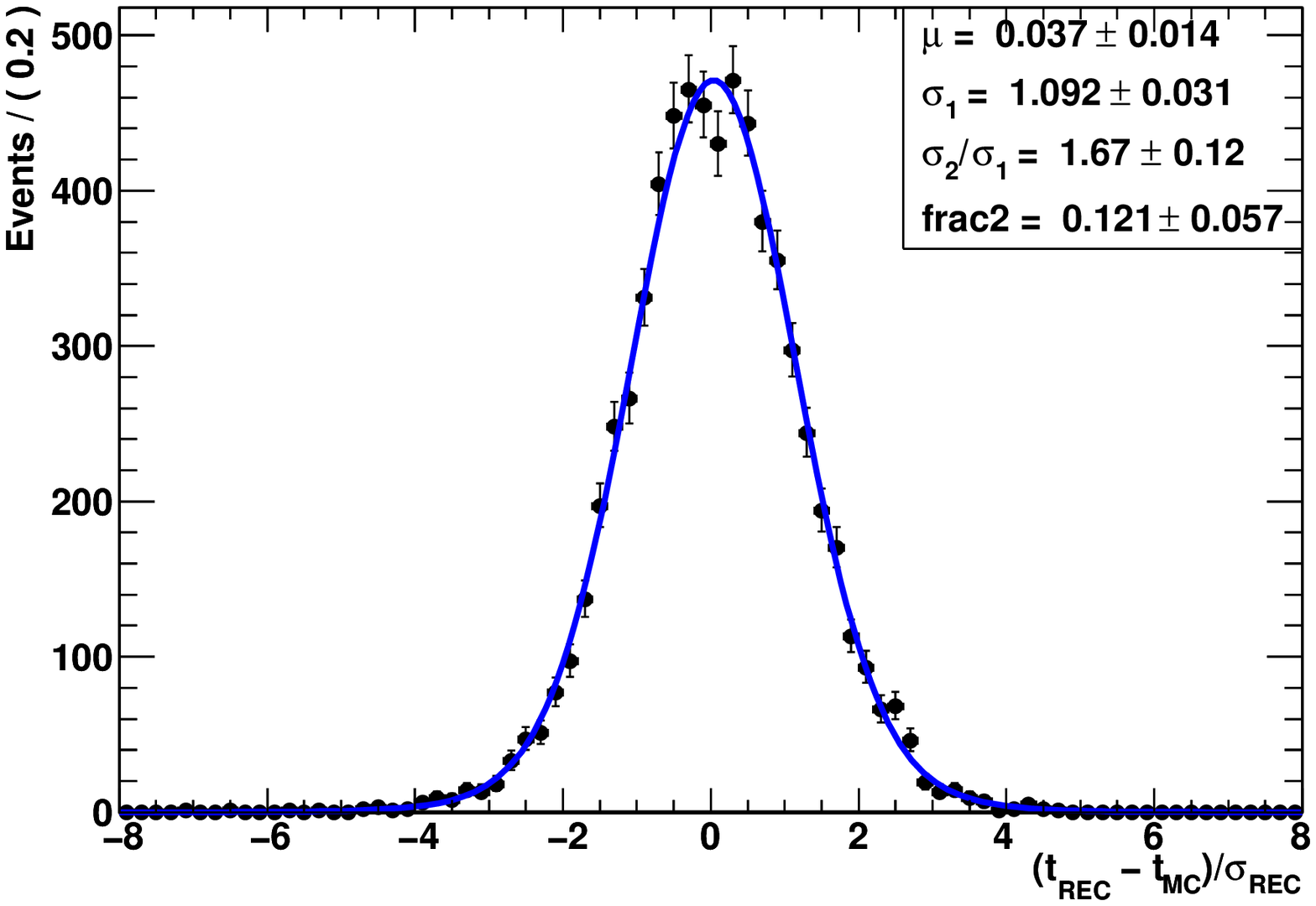}
\hbox to 0.45\textwidth{\hfil (a) $\sigma_t$  \hfil}%
\hbox to 0.45\textwidth{\hfil (b) $\frac{t^{\rm rec}-t^{\rm
MC}}{\sigma_t}$\hfil}\break \caption{\label{pull} Distributions of
(a) the proper time error estimate and (b) the proper time pull.}
\end{figure}

The time-dependent selection efficiency is shown in Fig.~\ref{acc}
for events before and after the trigger requirements. It is parameterized by the
acceptance function defined as:
\begin{equation}
\epsilon_t(t)= C \times \frac{(at)^n}{1+(at)^n},\label{eq-acc}
\end{equation}
where $C$ is the selection efficiency at large decay time, and $a$
and $n$ are two parameters which govern the proper time dependence at small
decay times.
\begin{figure}[htbp]
\center
\includegraphics[width=0.45\textwidth]{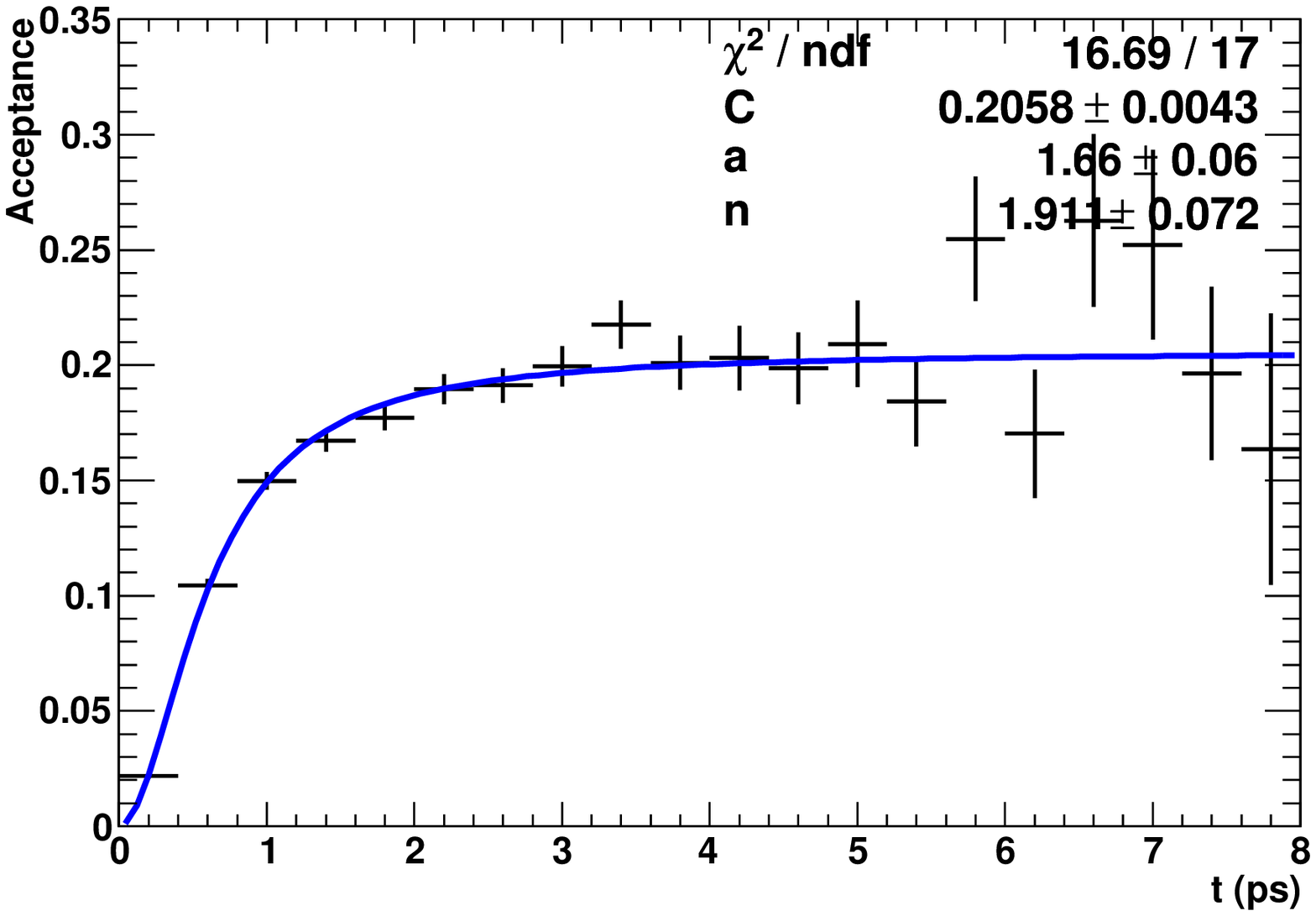}%
\includegraphics[width=0.45\textwidth]{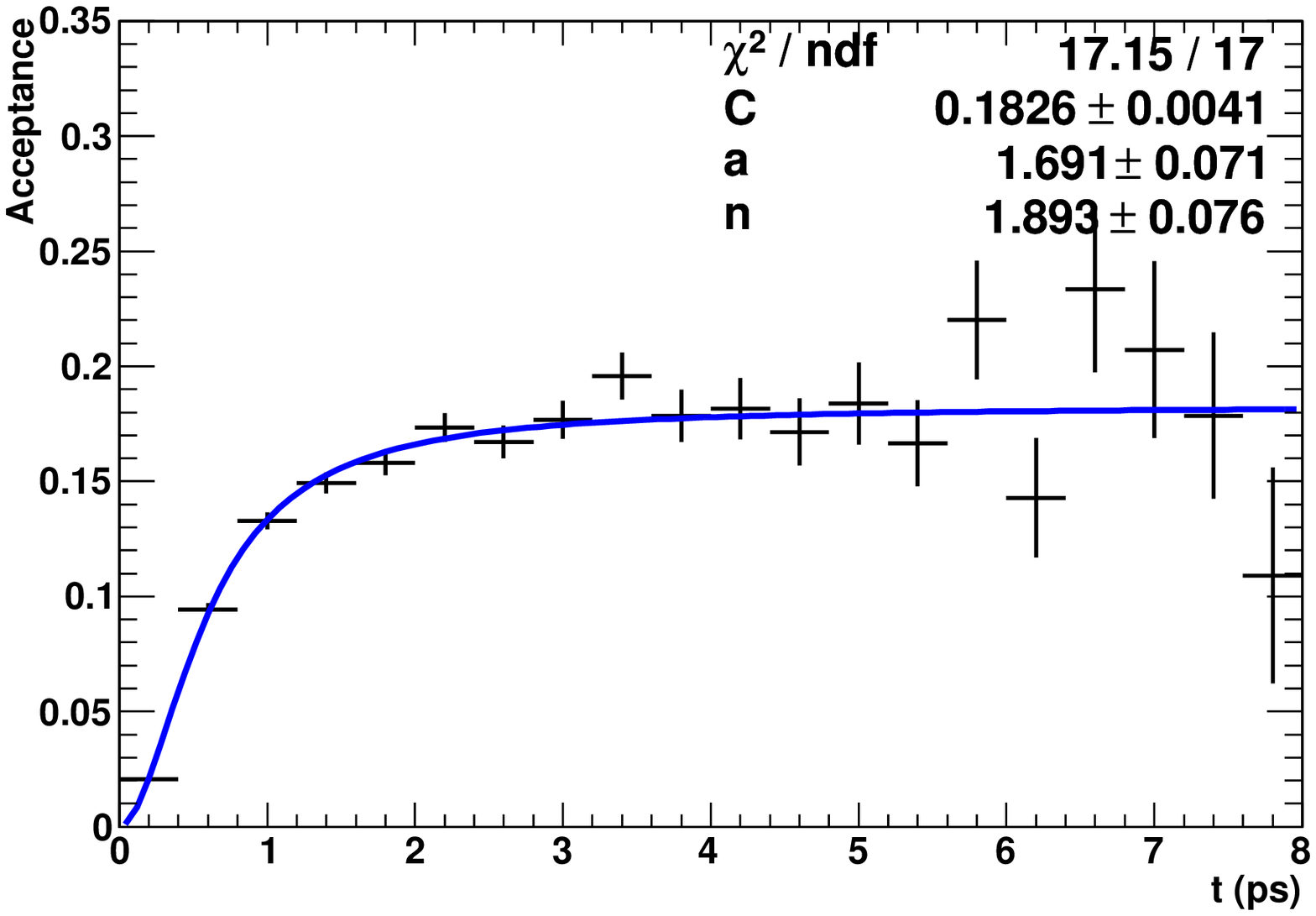}
\hbox to 0.45\textwidth{\hfil (a) untriggered \hfil}%
\hbox to 0.45\textwidth{\hfil (b) triggered \hfil}\break
\caption{\label{acc}The acceptance function as function of proper time.}
\end{figure}

\section{\boldmath Sensitivity of $\bs \to \Jpsi f_0$ Compared With $\Jpsi\phi$}
\label{sec:sens}
In order to perform this task a fast (Toy) Monte Carlo simulation
program has been used that is based on RooFit. We simulate 400 LHCb ''experiments" with 2 fb$^{-1}$
per sample. For each experiment the program generates events
taking into account the results obtained from the full MC simulation
and performs a fit according to a likelihood function which includes
the $\bs-\bsbar$ mixing variables. The sensitivity to $\phi_f$ is
taken as the root-mean-square width, $\sigma$, resulting from a Gaussian fit to the $\phi_f$
values from each simulation run.

\subsection{Likelihood function}
The time-dependent decay rates for initially pure $\bs$ or $\bsbar$
states in $\bar{b}\to \bar{c}c\bar{s}$ or $b\to c\bar{c}s$
quark-level transitions are given by the following expressions for
decay into CP eigenstates:
\begin{eqnarray}
\Gamma(\bs\to f_{CP}) &\propto& e^{-\Gamma_s
t}\left\{\cosh\frac{\Delta\Gamma_s
t}{2}-\eta_f\cos\phi_f\sinh\frac{\Delta\Gamma_s t}{2}+\eta_f
\sin\phi_f\sin(\Delta m_s t)\right\}\nonumber\\
\Gamma(\bsbar\to f_{CP}) &\propto& e^{-\Gamma_s
t}\left\{\cosh\frac{\Delta\Gamma_s
t}{2}-\eta_f\cos\phi_f\sinh\frac{\Delta\Gamma_s t}{2}-\eta_f
\sin\phi_f\sin(\Delta m_s t)\right\}\label{t-dep}
\end{eqnarray}
where $t$ is the proper time, $\eta_f$ is the CP eigenvalue of the state
$f_{CP}$, and $\Delta \Gamma_s$ is the lifetime difference between CP+ and
CP-- eigenstates. Direct CP violation is neglected. Note that for $J/\psi f_0$,
$\eta_f=-1$.

 The events are used to maximize a likelihood function ($\cal L$)
which is given by

\begin{equation}
{\cal L}=\prod_i^{N_{\rm obs}} P(m_i,t_i^{\rm rec}, q_i),
\end{equation}
with
\begin{equation}
P(m_i, t_i^{\rm rec}, q_i)=f_{\rm sig} P_m^{\rm sig}(m_i)P_t^{\rm
sig}(t_i^{\rm rec},q_i)+(1-f_{\rm sig})P_m^{\rm bkg}(m_i)P_t^{\rm
bkg}(t_i^{\rm rec}),
\end{equation}
where:
\begin{itemize}
\item $P_m^{\rm sig}(m_i)$ and $P_m^{\rm bkg}(m_i)$
are the probability
density functions (PDFs) describing the dependence on reconstructed mass $m_i$ for signal and background events;
\item $P_t^{\rm sig}(t_i^{\rm rec},q_i)$ is the PDF used to describe
the signal decay rates for the decay time $t_i^{\rm rec}$, which
depends on the tagging result at time t=0, $q_i$ ($q=+1$ if the signal meson is
tagged as $\bs$, $q=-1$ if it is tagged as $\bsbar$, or $q=0$ if no
tagging information);
\item $P_t^{\rm bkg}(t_i^{\rm rec})$ is the PDF describing the
background decay rates, which do not depend on the tagging
performance.
\item $f_{\rm sig}$ is the fraction of the signal in the fitting
region.
\end{itemize}

The likelihood function includes distinctive contributions from
the signal and the background. For both, the PDF is a production of
PDFs which model the invariant mass distribution and the
time-dependent decay rates. The PDF used for generating the $\bs$
mass spectrum consists of a double-Gaussian for the signal and a linear
function for the background. The PDF for the proper time is modeled
by a true decay function smeared by time resolution then multiplied
by the decay time acceptance function. From Eq. \ref{t-dep}, the
true time decay function for the signal can be expressed as:
\begin{equation}
R(t,q) \propto e^{-\Gamma_s t}\left\{\cosh\frac{\Delta\Gamma_s
t}{2}+\cos\phi_f\sinh\frac{\Delta\Gamma_s t}{2}- q D
\sin\phi_f\sin(\Delta m_s t)\right\}.\label{eq-sigt}
\end{equation}
The effect of the wrong-tag probability $w_{\rm tag}$ is included in
the dilution factor $D\equiv(1-2w_{\rm tag})$. We take this value as
fixed from studies of other modes such as $B_s^0\to D_s^-\pi^+$, and
use $w_{\rm tag}=0.334$ as our value. We take the wrong tag fractions
as the same for $\bs$ and $\bsbar$.

%\afterpage{\clearpage}
\subsection{Input to the fast Monte Carlo simulation}

The PDFs and their input parameters are shown in Table \ref{PDFs}.
The parameters of the signal are obtained from the distributions
shown in Fig.~\ref{mass}, \ref{dt} and \ref{acc} (b).  For the
parameters of physics and tagging performance, we use the same as
those used in $\bs \to \Jpsi \phi$ study \cite{jpsiphi}, as shown in
Table~\ref{input}. The PDF of the signal proper-time is shown
in Fig.~\ref{pdf-sig} for both $\phi_f=-0.0368$ (SM value) and
$\phi_f=-0.2$ rad, where the oscillations are apparent.
\begin{figure}[htb]
\center
\includegraphics[width=0.5\textwidth]{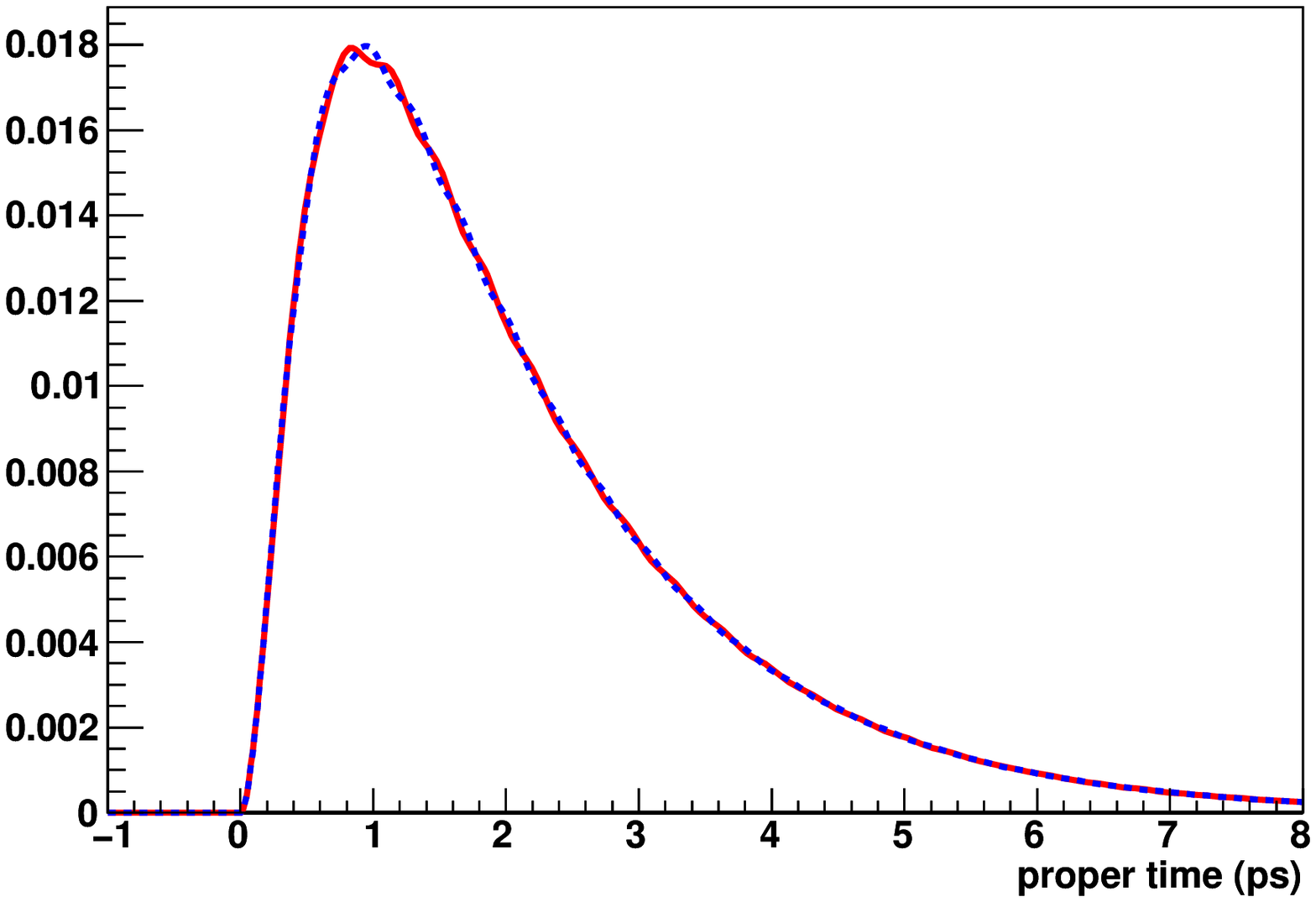}%
\includegraphics[width=0.5\textwidth]{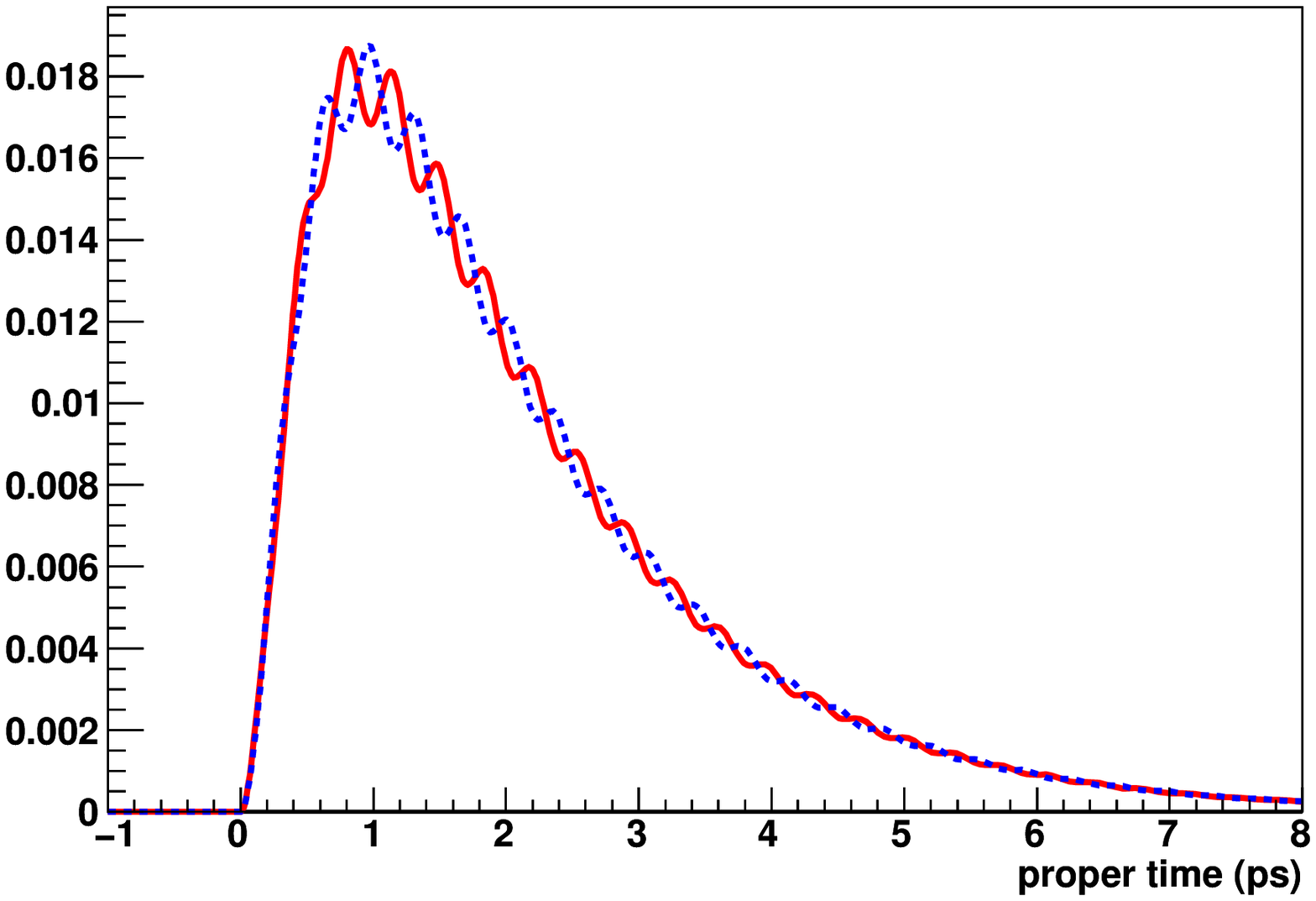}
\hbox to 0.5\textwidth{\hfil (a) $\phi_f=-0.0368$ rad \hfil}%
\hbox to 0.5\textwidth{\hfil (b) $\phi_f=-0.2$ rad \hfil}\break
\caption{\label{pdf-sig}The proper-time PDF of the signal for $\bs$
(solid line) and $\bsbar$ (dashed line).}
\end{figure}

The background's proper time distribution is obtained from $B_{u,d,s}\to \Jpsi X$ shown in
Fig.~\ref{t_bg}. The average proper time  is $1.93$ ps for the signal and 1.20 ps for the
$b\bar{b}$ background. We use $B/S=0.26$ for background from $b$ decays and $B/S=0.05$ for that from
prompt $\Jpsi$.\footnote{The tagging efficiency for the prompt $\Jpsi$ background is lower than
the signal and $b\bar{b}$ background \cite{jpsiphi}.}

\begin{table}[htb]
\center
\caption{\label{PDFs} The PDFs for the invariant mass and proper
time describing the signal and background.}
%\hspace{-1cm}
\begin{tabular}{c|c|c}\hline
&$P_m$ & $P_t$ \\\hline Signal&Double Gaussian ($2G$)&
\\
&$2G(m;m_0,\delta_m,\sigma_1,\sigma_2,f_2)$&$[R(t^{\rm MC},q)\otimes 2G(t^{\rm rec}-t^{\rm
MC};\mu,\sigma_1^t,\sigma_2^t,f_2^t)]\cdot \epsilon_t(t^{\rm rec};a,n)$ \\
&$m_0=5368.4$ MeV&$\mu=0.0008$ ps\\
&$\delta_m=-22$ MeV&$\sigma_1^t=0.0236$ ps\\
&$\sigma_1=16.8$ MeV&$\sigma_2^t=0.047$ ps\\
&$\sigma_2=33.2$ MeV&$f_2^t=0.36$\\
&$f_2=0.09$&$a=1.68$\\
&&$n=1.87$\\\hline

Background &First-order polynomial&$[e^\frac{-t^{\rm MC}}{\tau^{\rm bkg}}\otimes G(t^{\rm
rec}-t^{\rm MC};\mu,\sigma_1^t)]\cdot
\epsilon_t(t^{\rm rec};a,n)$\\
from $b$&&$\mu=0$\\
&&$\sigma_1^t=0.039$ ps\\
&&$\tau^{\rm bkg}=0.96$ ps\\
&& $a=4.4$\\
&&$n=2.9$\\
\hline

Background &First-order polynomial&$2G(t^{\rm rec}-t^{\rm
MC};\mu,\sigma_1^t,\sigma_2^t,f_2^t)\cdot
\epsilon_t(t^{\rm rec};a,n)$\\
from prompt &&$\mu=0$\\
$\Jpsi$&&$\sigma_1^t=0.11$ ps\\
&&$\sigma_2^t=2.4$ ps\\
&&$f_2^t=0.17$\\
&& $a=4.4$\\
&&$n=2.9$\\
\hline
\end{tabular}

\end{table}

\begin{table}[htb]
\center
\begin{tabular}{c|c}
Parameters& Input values \\\hline
$\Gamma_s$&$0.680$ps$^{-1}$\\
$\Delta \Gamma_s$& 0.049 ps$^{-1}$\\
$\Delta m_s$& $17.77$ps$^{-1}$\\
$\phi_f=-2\beta_s$&$-0.0368$ rad\\
$\varepsilon_{\rm tag}$& 0.564\\
$\omega_{\rm tag}$& 0.334\\\hline
\end{tabular}
\caption{\label{input} The input parameters of physics and tagging performance used for
fast Monte Carlo simulation.}
\end{table}

\begin{figure}[htb]
\center
\includegraphics[width=0.5\textwidth]{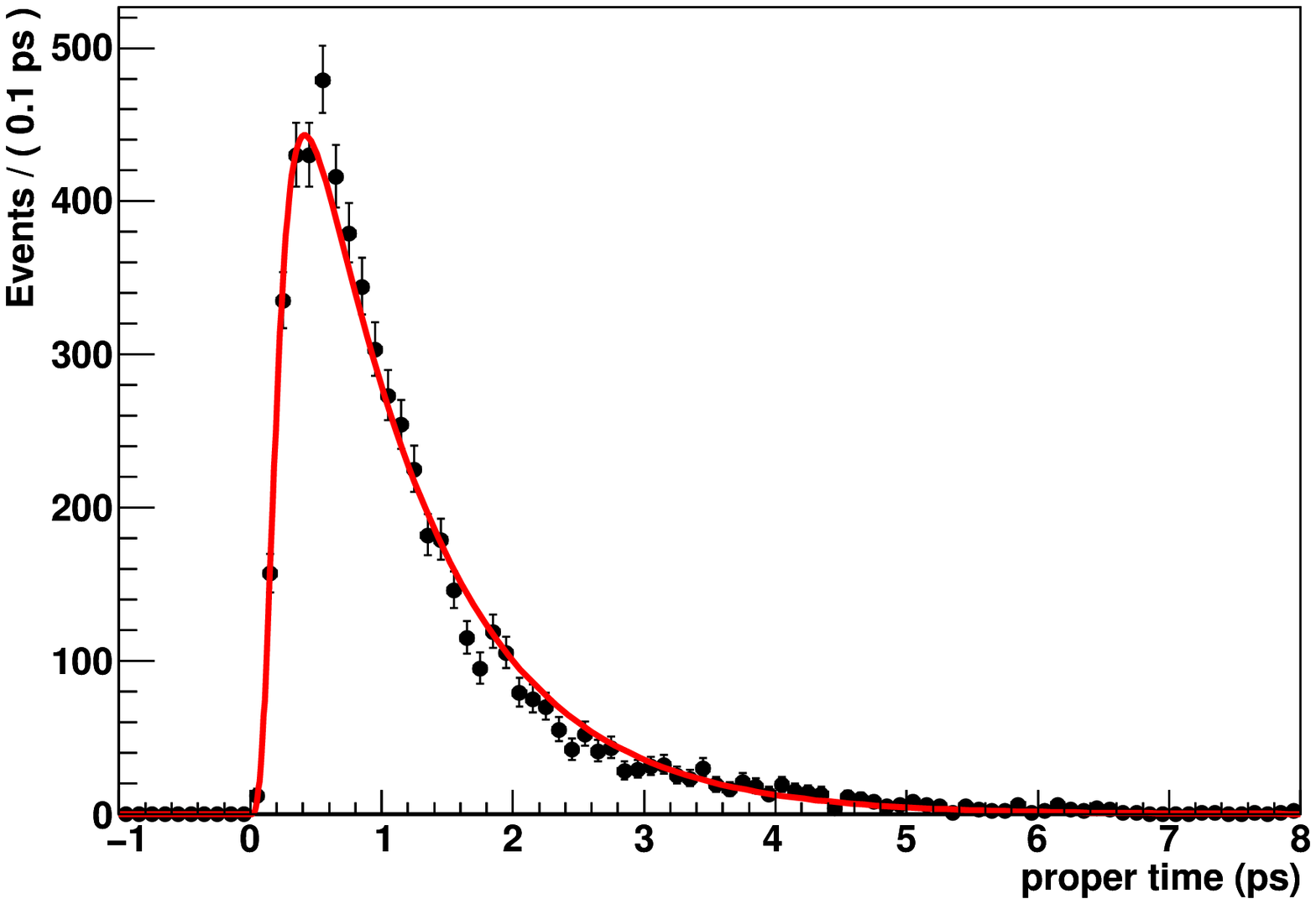}\break
\hbox to 0.5\textwidth{\hfil (a) background from $b$  \hfil} \break
\includegraphics[width=0.5\textwidth]{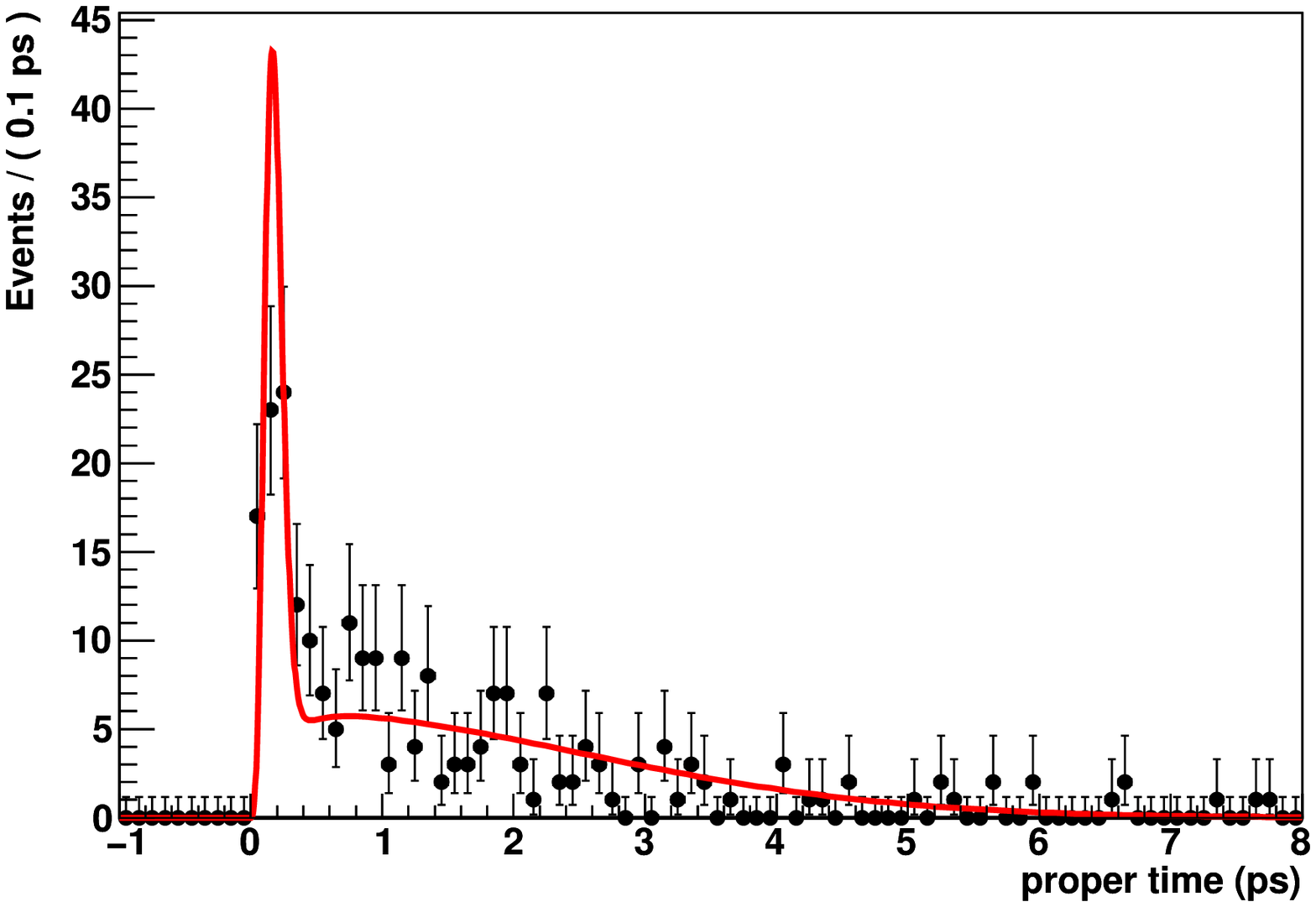}\break
\hbox to 0.5\textwidth{\hfil (b) background from prompt $\Jpsi$  \hfil} \break
 \caption{\label{t_bg}The
background proper time distributions from (a) from $B_{u,d,s}\to \Jpsi X$ and (b)
prompt $\Jpsi$.}
\end{figure}

%\subsection{Fit procedure}
%Taking into account the complexity of the likelihood function due to
%the large number of events and free parameters, the fit is performed
%in three successive steps:
%\begin{enumerate}
%\item Mass distribution fit:
%\end{enumerate}

\subsection{Fit Results}
 For this baseline study, we fit the Toy Monte Carlo only allowing $\phi_f$ and   $\Delta \Gamma$ to float, with
the other parameters fixed to their input values. Fig.~\ref{toy} shows the
distributions of mass and the proper time for the events tagged as $\bs$ and $\bsbar$ from one
experiment. The output of $\phi_f$ from the fits is shown in Fig.~\ref{sens} with signal only
and background included. The sensitivity of $\phi_f$ is estimated to be $(0.050\pm0.002)$ rad. We
also find that allowing the proper time acceptance parameters to float does not increase the uncertainty of $\phi_f$.

\begin{figure}[htbp]
\center
\includegraphics[width=1.0\textwidth]{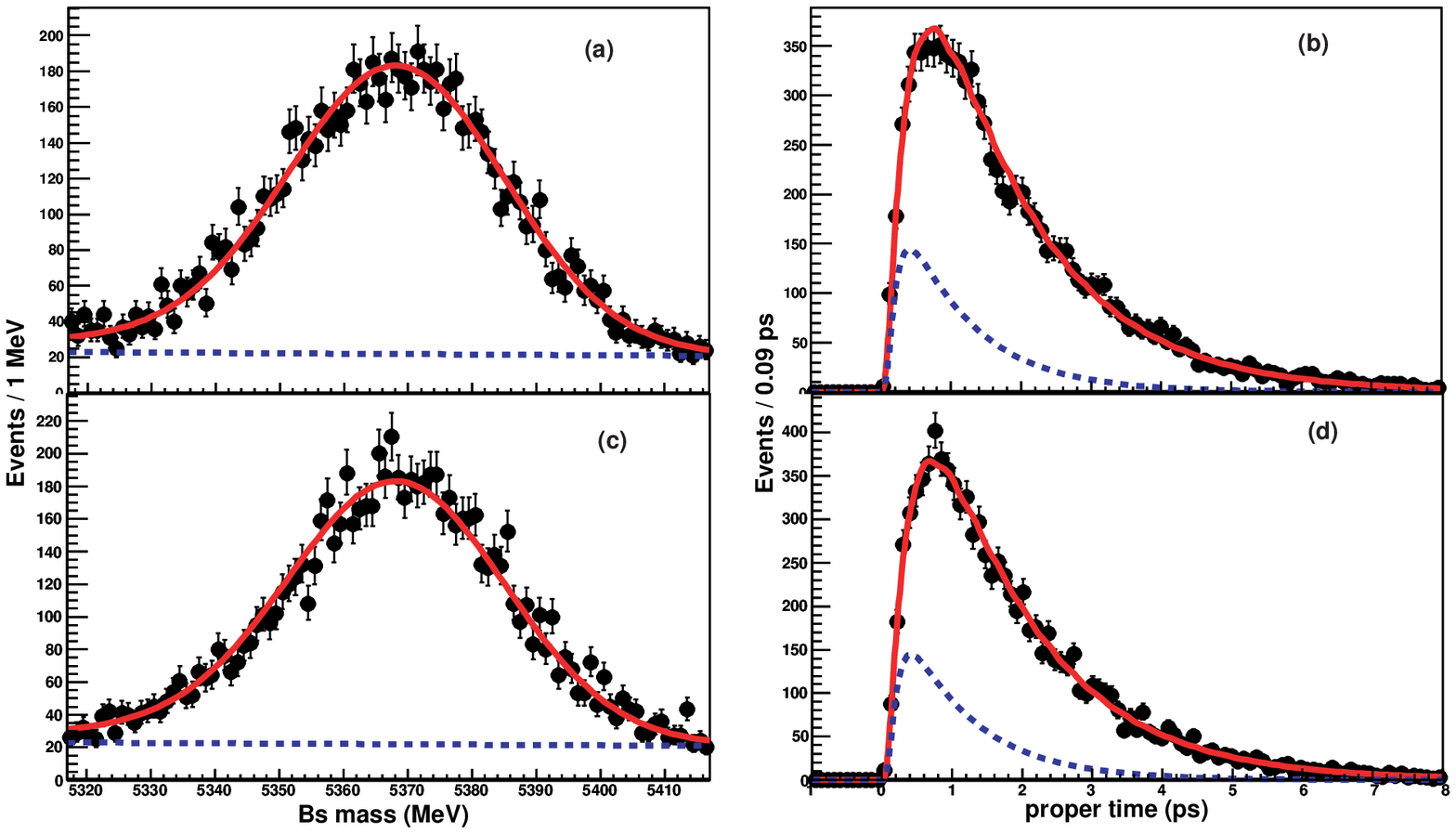}
\caption{\label{toy} (a) Invariant mass and  (b) proper-time distributions for the events
tagged as $\bs$, and (c) invariant mass and (d) proper-time distributions for those tagged
as $\bsbar$. The points with error bars are fast simulation data and the solid lines are fit
functions. The dashed lines show the background contribution.}
\end{figure}

\begin{figure}[htbp]
\center
\includegraphics[width=0.5\textwidth]{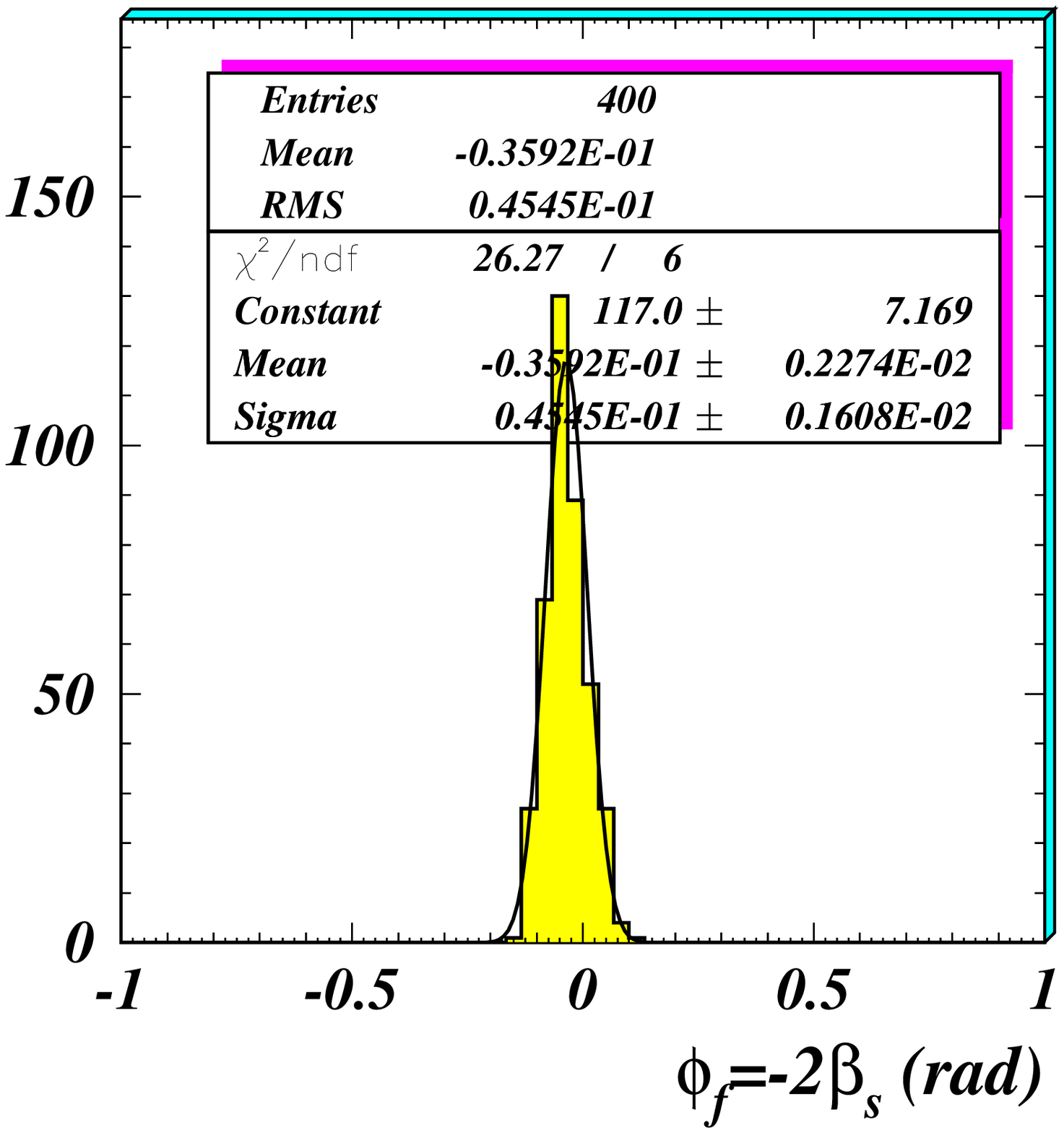}%
\includegraphics[width=0.5\textwidth]{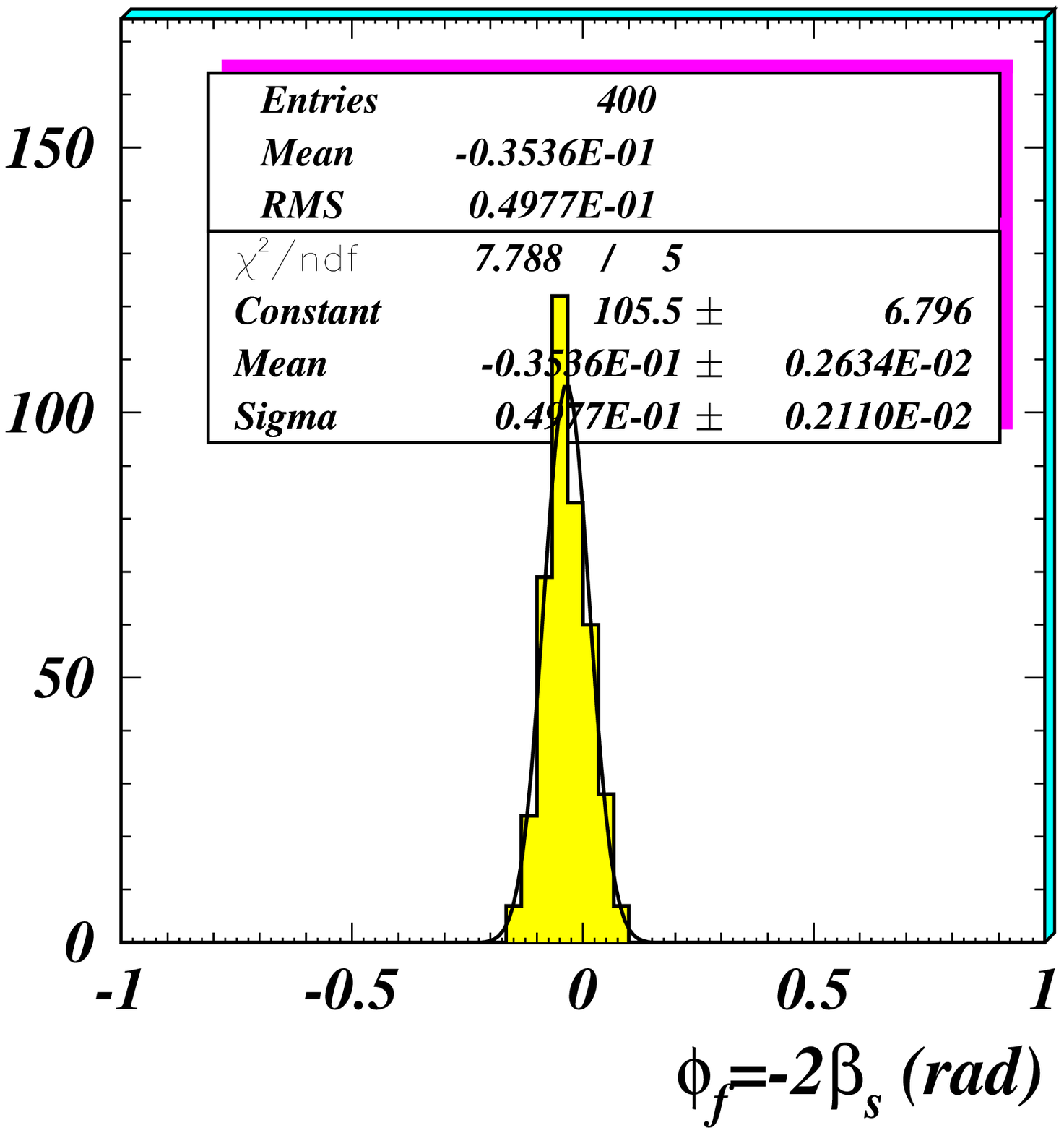}
\hbox to 0.5\textwidth{\hfil (a) signal only\hfil}%
\hbox to 0.5\textwidth{\hfil (b) signal+background \hfil}\break \caption{\label{sens}The
$\phi_f$ output from fitting to 400 toy MC.}
\end{figure}

\subsection{\boldmath Inclusion of $B_s\to J/\psi \eta'$ Events}
\label{subsec:etap}
We have already discussed the the contribution of the $ J/\psi \eta'$, $\eta'\to \rho^0\gamma$ events
as a source of background in the  $J/\psi f_0$ sample.  We also intend to reconstruct these events when we can find the photon from the $\eta'\to \rho^0\gamma$ decay and add them into our final sample (with reversed CP to $f_0$ events). In our simulation we use all photons found in the electromagnetic calorimeter and those that convert in
material in front of the magnet, provided that their $p_T$ is larger than 300 MeV/c. The detection efficiency for photons in the solid angle of the detector is about 25\%.
In 2 fb$^{-1}$ we estimate ~5000 of such fully reconstructed events before the trigger. A previous analysis of this mode
\cite{Jpsietap} concluded that an error in the measurement of  $\phi_f$ of $\pm$0.8 rad could be made with a 2 fb$^{-1}$ sample. Adding the two modes together would give an error in $\phi_f$ of $\pm$ 0.044 rad.

\section{Systematic Errors}
We have studied several sources of systematic error. Recall the outputs of our fit nominally are
$\phi_f$, $\Delta \Gamma$ and the time acceptance parameters ($a$ and $n$). We actually determine, however,
the product of the dilution $D$ times $\phi_f$ and use the value of $D$ determined from other measurements. Thus the systematic error on $D$ is fully correlated with the systematic error on $\phi_f$.

Now we will estimate the systematic error on $\omega_{tag}$. This parameter can be measured using a combination of other modes. One simple approach is to measure $\omega_{tag}$ using $B_s^0\to D_s^+\pi^-$. We expect differences in the value of $\omega_{tag}$ here and in the $J/\psi f_0$ mode because of different triggering in the hadronic and dimuon channels \cite{tagging}.
This difference can be estimated by simulation and checked using other modes. For example, we can use $B^0\to J/\psi K^{*0}$, $K^{*0}\to K^+\pi^-$ to separately measure opposite side tagging. As input to a first estimate we decided to see the difference in Monte Carlo between the $J/\psi f_0$ and $D_s^+\pi^-$ $B_s$ final states.

Our simulation of $B_s^0\to J/\psi f_0$ yields $\omega_{tag}=(0.326\pm 0.003)$, where
the error is purely statistical. Using exactly the same tagging code on the mode $B_s^0\to D_s^+\pi^-$, which of course we can and will use to measure $\omega_{tag}$,
gives a value of $(0.309\pm 0.003)$, again the error is statistical.  (We believe that we can measure $\omega_{tag}$ with excellent statistical precision, so we will ignore the statistical error.) The difference is
5.3\% in $\omega_{tag}$, or 10.6\% in $D$. We argue that to first order we can use the Monte Carlo simulation to make the 5.3\% correction to account for tagging differences between the two modes. Knowing that Monte Carlo is not perfect, we estimate that we will have $\approx 25$\% error or $\pm$1.3\% uncertainty on $\omega_{tag}$ after subjecting the Monte Carlo to various tests. Thus an systematic error of 2.6\% on $D$ should be achievable without superhuman efforts. We also expect that our belief in the Monte Carlo will change with time, hopefully improving.

Two other sources of error have been investigated thus far using
toy Monte Carlo. The general procedure is that we vary a parameter, or a shape parameterization, that has been fixed in the fit
by plus and minus the expected error, and then repeat the fit.  We record the difference of
$\phi_f$ output between this fit and the nominal fit. Then the distribution of the difference is used to set the error by
fitting to a single Gaussian. The mean of the Gaussian is taken as the systematic error for each particular
source. To check if the systematic errors depends on the central value of $\phi_f$, we use both
$\phi_f=-0.2$ and $\phi_f=-0.736$ for generation. We notice that the systematic error
does depend on $\phi_f$ and $\sigma_{\phi_f}/\phi_f$ is approximately constant.
We plan to measure the time resolution in real data from prompt $J/\psi\to \mu^+\mu^-$, where we add two other tracks from the primary vertex consistent with the $f_0$ mass. We have seen that the pull distribution of time resolution for the signal and the prompt events, so defined,  are identical.
Our results are listed in Table~\ref{tab:sys}.
\begin{table}[htb]
\center
\caption{\label{tab:sys} Systematic error sources on $\phi_f$}
\begin{tabular}{lcc}
Source & Value & $\sigma_{\phi_f}/\phi_f$ (\%) \\\hline
Time resolution & 34 fs varied by $\pm$5\% & 2.5\\
Misstag rate & 0.334$\pm$0.004 & 2.6\\
$B_s$ lifetime & 1.470$\pm$0.027 fs & 3.0\\\hline
Total & & 4.7\\\hline
\end{tabular}

\end{table}

These errors are of comparable sizes. They
will improve with increased statistics especially in the control channels. None of
the systematic errors changes our estimation of statistics necessary for a significant effect, but merely
scales the value of $\phi_f$. We are continuing these preliminary studies.

Other more global sources of systematic errors due to $B_s$ production and
$B_s$ decay rates are discussed in the Conclusions.

\section{Conclusions}
We expect 26,100 $B_s\to J/\psi f_0$, $f_0\to \pi^+\pi^-$ signal events in 2 fb$^{-1}$ of accumulated LHCb data.
Based on branching fraction predictions of resonant $B_s^0\to J/\psi f_0(980)$,
$f_0(980)\to\pi^+\pi^-$ and non-resonant $B_s^0\to J/\psi\pi^+\pi^-$, where the two pions are
S-wave \cite{stone}, we predict an error on the measurement of the CP violating parameter
$\phi_f$ of $\pm0.050$ rad. Adding in the $J/\psi\eta'$, $\eta'\to\rho\gamma$ final state that
we need to measure to estimate backgrounds, reduces the error on $\phi_f$ to $\pm$0.044 rad.
This is larger than the estimate using the $J/\psi \phi$ final state of $\pm$0.03 rad, however
the latter estimate does not consider the effect of a $K^+K^-$ S-wave. Initial indications are
that taking the S-wave into account will increase the error by less than 15\% \cite{S-wave-sens}. Use of the $J/\psi$
plus scalar or pseudoscalar CP eigenstates removes the need for a complicated angular analysis
and should provide, at minimum, a crucial check on the vector-vector result. Both methods have
a large $\approx \pm$25\% systematic uncertainty on the predicted sensitivity due to uncertainties in the production cross-section
and $B_s$ branching ratios. In addition, the estimate based here assumes that relative yield of the
$\pi^+\pi^-$ S-wave in the $f_0$ mass region is 25\% that of $K^+K^-$ in the $\phi$ mass region.
This introduces a considerable uncertainty of about $\pm$40\% in the relative rates and $\pm$20\%
in the sensitivity, giving an overall uncertainty in of $\pm$32\% in the error in $\phi_f$.

\section{Acknowledgments}
We thank the U. S. National Science Foundation for support. We also thank our LHCb colleagues for useful discussions
and review of this paper.
\afterpage{\clearpage}


\begin{thebibliography}{99}
\bibitem{W-exchange}
There is the possibility of an additional decay diagram involving $W$-exchange, where we end up with a $u \overline{u}$ system with the $J/\psi$ being formed from additional gluons. If this diagram is significant and if the $f_0$ has a relatively large $u \overline{u}$ piece in its wave function, it could be a source of an additional phase. Current estimates of the light quark content in the $f_0$ wavefunction, however, are small, see Ref. \cite{stone} and references therein.
\bibitem{constraint}
We have not used a constraint on $J/\psi$ mass to improve the mass resolution because of the possibilites of introducing biases. This may be implemented in the future.
\end{thebibliography}
\end{document}